\documentclass[a4paper,11pt]{article}
\usepackage{jcappub}    
\usepackage{lineno}
\pdfoutput=1

\usepackage{graphicx}   
\usepackage{tabularx}
\usepackage{dcolumn}    
\usepackage{bm}         
\usepackage{hyperref}   
\usepackage[utf8]{inputenc}
\usepackage{amsmath,amssymb,amsthm}
\usepackage{orcidlink}
\usepackage{float}

\newcommand{\AddrLUPM}{Laboratoire Univers et Particules de Montpellier (LUPM), \\ CC 72, Place Eugène Bataillon, 34095 Montpellier - Cedex 5, France}
\newcommand{\AddrKIT}{Institut f\"ur Astroteilchenphysik, Karlsruher Institut f\"ur Technologie (KIT), Hermann-von-Helmholtz-Platz 1, 76344 Eggenstein-Leopoldshafen, Germany}

\title{\boldmath{Neutrino mass limits and decaying dark matter: background evolution versus perturbations}} 

\author[a]{Thomas Montandon\orcidlink{/0000-0002-5536-6953}}
\author[a]{Vivian Poulin\orcidlink{0000-0002-9117-5257}}
\author[b]{Thomas Rink\orcidlink{0000-0002-9293-1106}}
\author[b]{Thomas Schwetz\orcidlink{0000-0001-7091-1764}}

\affiliation[a]{\AddrLUPM}
\affiliation[b]{\AddrKIT}

\emailAdd{thomas.montandon@umontpellier.fr, vivian.poulin@umontpellier.fr, thomas.rink@kit.edu, schwetz@kit.edu}

\date{\today}   

\abstract{
We revisit cosmological neutrino mass bounds when a fraction of dark matter is allowed to decay to massless dark radiation. By compensating the late-time increase in the matter density induced by neutrinos becoming non-relativistic, decaying dark matter (DDM) can render datasets solely sensitive to the background density effectively insensitive to neutrino masses. Using data from baryonic acoustic oscillations (BAO) and Type~Ia supernovae together with a distance prior from the cosmic microwave background (CMB), we find that neutrino masses as large as ${\cal O}(1\,\mathrm{eV})$ are allowed without degrading the fit. Moreover, the combination of BAO data with the CMB distance prior yields a preference for a non-zero DDM fraction, and alleviates the need for dynamical dark energy with phantom crossing. However, the degeneracy introduced by DDM is decisively broken once perturbation observables are included. Incorporating the full \textit{Planck} CMB likelihood, and in particular CMB lensing, restores strong constraints on the neutrino mass in the DDM scenario, $\sum m_\nu \lesssim 0.079\,\mathrm{eV}$. In contrast, neutrino mass constraints in a smooth dark energy model described by the Chevallier-Polarski-Linder parametrization become merely $\sim 25\%$ stronger compared to background-only analyses. Our results highlight the essential role of structure-growth measurements in assessing extensions of the dark sector and to obtain robust cosmological neutrino mass bounds.
}

\begin{document}

\maketitle
\flushbottom


\section{Introduction}


The origin and values of neutrino masses and the fundamental properties of the dark sector remain some of the most persistent open questions in contemporary cosmology and particle physics. While terrestrial experiments have established that at least two neutrino mass eigenstates are non-zero \cite{Fukuda:1998mi,Ahmad:2002jz,Eguchi:2002dm,Esteban:2024eli}, cosmological observations currently provide the most stringent upper bounds on the absolute mass scale. Within the standard $\Lambda$ cold dark matter ($\Lambda$CDM) framework, where dark energy (DE) is described by a cosmological constant ($\Lambda$) and dark matter (DM) by a cold matter component solely coupled to gravity, large-scale structure (LSS) and cosmic microwave background (CMB) data jointly limit the sum of neutrino masses to the sub-0.1\,eV regime \cite{Planck:2018vyg,Elbers:2025vlz,ACT:2025nti,SPT-3G:2025bzu}. In fact, recent measurements of baryon acoustic oscillations (BAO) from the Dark Energy Spectroscopic Instrument (DESI) collaboration~\cite{DESI:2025zgx}, combined with CMB observations from the {\it Planck} satellite \cite{Planck2018I}, Atacama Cosmology Telescope (ACT)~\cite{AtacamaCosmologyTelescope:2025blo} and South Pole Telescope (SPT) \cite{SPT-3G:2025bzu}, have set the stringent bound
$$
\sum m_\nu < 0.048~{\rm eV}~(95\%\, {\rm C.L.})\, ,
$$
leading to an apparent tension with the terrestrial lower bound of $\sum m_\nu \gtrsim 0.06 ~{\rm eV}$ in the ``normal ordering'' scenario \cite{Gariazzo:2023joe,Esteban:2024eli}. We refer to Refs.~\cite{Craig:2024tky,Bertolez-Martinez:2024wez,Loverde:2024nfi,Naredo-Tuero:2024sgf,Jiang:2024viw,Lynch:2025ine,Graham:2025dqn,Giare:2025ath,Gorbunov:2026sly} for discussions on the origin and robustness of this tension.

It is well known that these constraints rely on assumptions about the behavior of DM and DE at late times. In fact, it has been shown that a simple extension to the $\Lambda$CDM model where the DE equation of state parameter, $w$, is allowed to vary over time as described by the Chevallier-Polarski-Linder (CPL) parametrization \cite{Chevallier:2000qy,Linder:2002et}, $w(a)=w_0+w_a(1-a)$ with $a$ denoting the cosmic scale factor normalized to $a=1$ today, can relax neutrino mass constraints to the ballpark of $\sim{\cal O}(0.15-0.20) \, {\rm eV}$ depending on the exact data combination \cite{Elbers:2024sha,Elbers:2025vlz}. Consequently, the tension with terrestrial experiments is reduced regardless of the neutrino mass ordering. Interestingly, the combination of CMB, DESI BAO and SN1a data exhibits a preference for dynamical dark energy (DDE) over $\Lambda$ at the $\sim3-4\sigma$ level \cite{DESI:2025zgx}. 

Yet, the favored DDE model is arguably ``exotic'', as it must cross the phantom-divide \cite{DESI:2024aqx,DESI:2025fii}, i.e.\ $w(a\ll 1)<-1 \to w(a\sim1)>-1$, hereby requiring a non-minimally coupled scalar field \cite{DESI:2024kob,DESI:2025fii,Wolf:2024stt,Teixeira:2024qmw,Cataneo:2025vae,Smith:2025grk,Khoury:2025txd}. Moreover, its average effect when integrated up to the last scattering surface must be compatible with a cosmological constant ($\langle w\rangle\simeq -1$), in order not to spoil the geometric degeneracy in the CMB imposed by the tight measurement of the acoustic scale, such that currently the DDE model favored by the data is the rather ad-hoc ``mirage dark energy'' model \cite{Linder:2007ka,DESI:2024kob,DESI:2025fii}. As stage IV LSS surveys are beginning to map cosmic expansion and structure growth with unprecedented accuracy \cite{LSST:2008ijt,Euclid:2025overview}, it becomes increasingly important to assess how non-standard dark sector physics may alter or relax cosmological neutrino mass bounds, and to understand whether alternative new physics scenarios may show similar preference over $\Lambda$CDM.

In this paper, we revisit neutrino mass bounds in different beyond-$\Lambda$CDM scenarios, when the assumption of stability of CDM over cosmological timescales is relaxed \cite{Audren:2014bca,Poulin:2016nat,FrancoAbellan:2020xnr,FrancoAbellan:2021sxk,Nygaard:2020sow,Simon:2022ftd,Holm:2022kkd,Bucko:2022kss,Alvi:2022aam,Holm:2023uwa,Montandon:2025xpd}. The main rationale is that, as it has been argued in the literature \cite{Poulin:2016nat,Lynch:2025ine}, DM decays to a light dark sector (typically a ``dark radiation'' particle that is not charged under the standard model) can be thought of as having the opposite effect that massive neutrinos have on the total energy density of the universe. Indeed, neutrinos become non-relativistic when their average momentum drops below their mass, $\langle p_\nu\rangle=3T_\nu\sim m_\nu$. This transition causes the expansion rate $H\equiv \dot{a}/a$ to increase slightly (compared to a massless neutrino universe) due to the additional matter component ``created'' at late times. Conversely, {\it decaying dark matter} (DDM) with massless decay products experiences a transition from a matter component to a radiation one when the Hubble rate is of the order of its decay rate $\Gamma_{\rm DDM} \sim H(a)$. If the decay rate and abundance of the DDM component are adjusted correspondingly, it is possible to compensate the effect of the neutrino non-relativistic transition on the expansion rate of the universe \cite{Poulin:2016nat,Lynch:2025ine}. 

In fact, we show here that for datasets that solely rely on cosmological background information (CMB and BAO scales in particular) there is a nearly exact degeneracy between neutrino mass and DDM and no meaningful bound can be placed on $\sum m_\nu$ based on these observables. Interestingly, at the background level, we show that DDM performs significantly better than DDE in relaxing the neutrino mass bounds, while having a weak preference over $\Lambda$CDM, hereby offering an alternative explanation to the DESI results. We further show that the combination of DDM and DDE can significantly alter the $\{w_0,w_a\}$ posteriors, reducing the need for phantom crossing. The apparent phantom behavior is mimicked by the decaying component, while a value of $w_0>-1$ remains required to provide a good fit to Type~Ia supernovae (SN1a) data.

We then investigate constraints when including the full {\it Planck} CMB data, beyond the ``background-only'' information. While neutrino masses and DDM have opposite effects at the background level, their effects at the level of matter perturbations add up \cite{Poulin:2016nat}. The inclusion of probes of the growth of structures such as CMB lensing thus allows to break the degeneracy between DDM and neutrino masses, yielding constraints on the sum of neutrino masses of $\sum m_\nu < 0.079$\,eV, only $\sim 10\%$ weaker than those obtained assuming standard $\Lambda$CDM. On the other hand, at the perturbation level, DDE described with the CPL parametrization, assuming a sound speed of perturbations $c_s^2=1$, has little impact on the growth of structures, and thus can relax neutrino mass bounds in an analysis that includes the full datasets to a similar extent as for the background-only analysis ($\sum m_\nu \lesssim 0.15$\,eV).

This suggests a novel model-building avenue: models with similar background effects as DDM, but leaving the growth of perturbations unaltered compared to $\Lambda$CDM, may perform better than the models studied in this work. In fact, models that couple dark matter and dark energy \cite{Wolf:2024stt,Teixeira:2024qmw,Cataneo:2025vae,Smith:2025grk,Khoury:2025txd} appear to have those features. For a detailed discussion on dark forces and their impact on structure growth see also \cite{Costa:2025kwt}.

Our paper is structured as follows: In Sec.~\ref{sec:models}, we discuss the DDM and DDE models and their interplay with neutrino masses at the background level.  In Sec.~\ref{sec:background}, we perform a ``background-only'' analysis of CMB and DESI BAO distance observables, to investigate the extent to which constraints on neutrino masses coming from their background effect alone can be relaxed. We also test how the combination of DDM and DDE alter constraints to the equation of state of dark energy parameters. To guarantee the robustness of our results we compare constraints derived in a Bayesian and in a Frequentist framework. In Sec.~\ref{sec:cmb}, we derive constraints using the full power of CMB data and highlight in particular the role played by CMB lensing. We finally summarize our results and conclude in Sec.~\ref{sec:conclusions}.


\section{Extended dark sector and neutrino masses}\label{sec:models}


In this section we describe the cosmological framework considered throughout this work, focusing on the CPL parametrization of the equation of state of DE, the background evolution of DDM, and their interplay with massive neutrinos.


\subsection{Dynamical dark energy}

A DDE component whose equation of state varies over cosmic time is conventionally described by the Chevallier--Polarski--Linder (CPL) parametrization \cite{Chevallier:2000qy,Linder:2002et}
\begin{equation}
w(a) = w_0 + w_a (1-a)\, ,
\label{eq:CPL}
\end{equation}
where $a$ stands for the scale factor. It provides a flexible two-parameter description capturing a wide range of quintessence- or phantom-like behaviors. The corresponding dark energy density evolves as
\begin{equation}
\rho_{\rm DE}(a) = \rho_{\rm DE,0}\,
a^{-3(1+w_0+w_a)}\,
\exp\!\Big[-3w_a(1-a)\Big]\, .
\label{eq:CPL_rho}
\end{equation}
In the limit $(w_0,w_a)=(-1,0)$, Eq.~\eqref{eq:CPL_rho} reduces to a cosmological constant. 
As is customary in {\tt CLASS}, dark energy perturbations are treated within the standard parametrized post-Friedmann (PPF) framework \cite{Fang:2008sn}, assuming a sound speed $c_s^2=1$, such that dark energy remains smooth on sub-horizon scales. 

Recent BAO measurements from the DESI collaboration (DR2) \cite{DESI:2025zgx} have provided the most precise low-redshift angular distance constraints to date, enabling a stringent test of DDE models. When interpreted within the CPL parametrization, DESI data combined with CMB and Type~Ia supernovae exhibit a statistically significant preference for a time-varying equation of state, typically at the ${\sim}3$--$4\sigma$ level, with $w_0 > -1$ and $w_a < 0$. This apparent evolution of dark energy is closely tied to the emerging tension in the matter density parameter $\Omega_m$, where DESI favors slightly lower values than those inferred from \textit{Planck} $\Lambda$CDM and significantly smaller than those measured by SN1a compilations, Pantheon+ \cite{Brout:2022vxf}, DESY5 \cite{DES:2024jxu} or Union3 \cite{Rubin:2023jdq}. The CPL model attempts to reconcile these discrepant preferred densities by allowing $w(a)$ to evolve, effectively adjusting the late-time expansion history. While the physical interpretation remains under investigation, these results have intensified interest in alternative dark sector models capable of reproducing similar signatures without invoking phantom crossing. We show such an example with the decaying cold dark matter scenario.


\subsection{Decaying cold dark matter}

We consider a DDM scenario in which a fraction of the cold dark matter component is unstable and decays into an invisible relativistic species (``dark radiation''), with no direct coupling to the Standard Model. The decay is characterized by a constant decay rate
\begin{equation}
\Gamma \equiv \frac{1}{\tau_{\rm DDM}}\, ,
\end{equation}
where $\tau_{\rm DDM}$ is the lifetime of the unstable component. The background energy densities of the DDM, $\rho_{\rm DDM}$, and dark radiation components, ${\rho}_{\rm DR}$, obey \cite{Poulin:2016nat}

\begin{align}
\dot{\rho}_{\rm DDM} + 3H \rho_{\rm DDM} &= - \Gamma\, \rho_{\rm DDM}\, , 
\label{eq:dcdm_rho}\\[4pt]
\dot{\rho}_{\rm DR} + 4H \rho_{\rm DR} &=  \Gamma\, \rho_{\rm DDM}\, ,
\label{eq:dr_rho}
\end{align}
where an overdot denotes a derivative with respect to cosmic time and $H=\dot a/a$ stands for the Hubble expansion rate.  
Equation~\eqref{eq:dcdm_rho} shows that the DDM energy density redshifts as non-relativistic matter but decays exponentially with lifetime $\tau_{\rm DDM}$, while Eq.~\eqref{eq:dr_rho} describes the corresponding energy injection into a relativistic component with equation of state parameter $w_{\rm DR}=1/3$. The total Hubble expansion rate thus exhibits a characteristic late-time reduction in the effective matter density, opposite to the behavior induced by massive neutrinos. This opposite evolution underlies the strong degeneracy between DDM and neutrino masses at the background level.

We parameterize the initial abundance of the decaying component as
\begin{equation}
f_{\rm DDM} \equiv \frac{\rho^{\rm ini}_{\rm DDM}}{\rho^{\rm ini}_{\rm DDM} + \rho_{\rm CDM}^{\rm stable}}\, ,
\end{equation}
and treat both $f_{\rm DDM}$ and $\Gamma_{\rm DDM}$ as free parameters. The standard $\Lambda$CDM limit is recovered for $\Gamma_{\rm DDM} \to 0$ or $f_{\rm DDM} \to 0$. Within this framework the cosmology is independent of the specific particle physics realization of the DM decay, in particular also of the DM mass. A more general class of scenarios converting DM into dark radiation has been discussed in \cite{Bringmann:2018jpr}.
Let us mention that a similar behavior could arise in models of neutrino decays to massless dark particles~\cite{Doroshkevich:1989,Doroshkevich:1989bf,Chacko:2019nej,Barenboim:2020vrr,Escudero:2020ped,Chacko:2020hmh,FrancoAbellan:2021hdb,FrancoAbellan:2026ori}. Such models lead to similar background phenomenology as DDM with massive neutrinos, but a very different impact on the matter clustering \cite{Chacko:2019nej,FrancoAbellan:2021hdb}.


\subsection{Massive neutrinos and degeneracies with the dark sector}

Massive neutrinos leave imprints on both the background expansion and the growth of cosmic structures. Standard model neutrinos have decoupled from the thermal bath at about $T\sim 1~{\rm MeV}$ and remain in a frozen Fermi-Dirac distribution, simply cooling down under the effect of the Universe's expansion \cite{Lesgourgues:2018ncw}. As their average momentum reaches the order of magnitude of their mass, $\langle p_\nu\rangle=3T_\nu\sim m_\nu$, their equation of state transitions from that of radiation $w_\nu \simeq 1/3$ to that of matter $w_\nu \simeq 0$, effectively increasing the late-time matter density following the approximate relation \cite{Lesgourgues:2006nd}
\begin{equation}
\label{eq:omega_nu}
    \Omega_\nu h^2 = \frac{\sum m_\nu}{93.14~{\rm eV}}\, .
\end{equation}
This alters the distance to recombination and therefore the acoustic scale measured by the CMB. 
This effect allows the combination of BAO and CMB data to break the degeneracy between the expansion rate today $H_0$ and the matter density fraction $\Omega_m$ \cite{Lesgourgues:2006nd,Lesgourgues:2018ncw,Archidiacono:2016lnv,Bertolez-Martinez:2024wez,Naredo-Tuero:2024sgf,Loverde:2024nfi,Elbers:2025vlz}. In addition to this background effect, neutrino free-streaming suppresses the growth of structure on scales $k \gtrsim 0.01\,h\,{\rm Mpc}^{-1}$, leaving characteristic signatures in CMB lensing, galaxy clustering, and weak-lensing observables.

Recent cosmological analyses combining \textit{Planck}, ACT, SPT, and DESI DR2 BAO have pushed the upper bound on the sum of neutrino masses into the sub-0.1\,eV regime, with constraints as stringent as $\sum m_\nu < 0.05$--$0.07\,{\rm eV}$ (95\% C.L.) for the $\Lambda$CDM model \cite{ACT:2025nti,SPT-3G:2025bzu,Elbers:2025vlz}. 
These limits are in mild tension with the terrestrial lower bound for normal ordering, $\sum m_\nu \gtrsim 0.06\,{\rm eV}$, motivating scrutiny of the underlying cosmological assumptions.
In fact, some analyses even report an unphysical preference for \emph{negative} values of the inferred neutrino mass when the prior is extended below zero \cite{Craig:2024tky,Elbers:2024sha,Naredo-Tuero:2024sgf,Graham:2025dqn,Elbers:2025vlz}.   
The preference for negative neutrino mass can ultimately be traced back to two effects: i) the preference of recent DESI data for lower $\Omega_m$ than inferred under \textit{Planck} $\Lambda$CDM and by SN1a data (a trend that CPL accommodates by shifting $w_0 > -1$ and $w_a < 0$) and ii) the infamous `$A_{\rm lens}$' anomaly in older version of the \textit{Planck} high-$\ell$ likelihood (plik PR3 \cite{Planck:2018vyg}), that manifests as a higher level of smoothing of the acoustic peaks than predicted under $\Lambda$CDM and measured by the lensing power spectrum.

Allowing for a time-varying DE equation of state through the CPL parametrization significantly modifies these bounds. Since $w_0$ and $w_a$ alter the late-time expansion history, CPL models can partially compensate the background effect of massive neutrinos, thereby weakening the geometric constraints from CMB+BAO. Indeed, when $w(a)$ is allowed to deviate from $-1$, the neutrino mass upper limit degrades to $\sum m_\nu \sim {\cal O}(0.15-0.20)\,{\rm eV}$ depending on the dataset combination, reflecting a strong degeneracy between neutrino mass and the integrated evolution of $w(a)$. Moreover, the preference for negative neutrino masses disappears \cite{Elbers:2024sha,Naredo-Tuero:2024sgf,Elbers:2025vlz}.

\begin{figure}[h]
    \centering
    \includegraphics[width=0.49\linewidth]{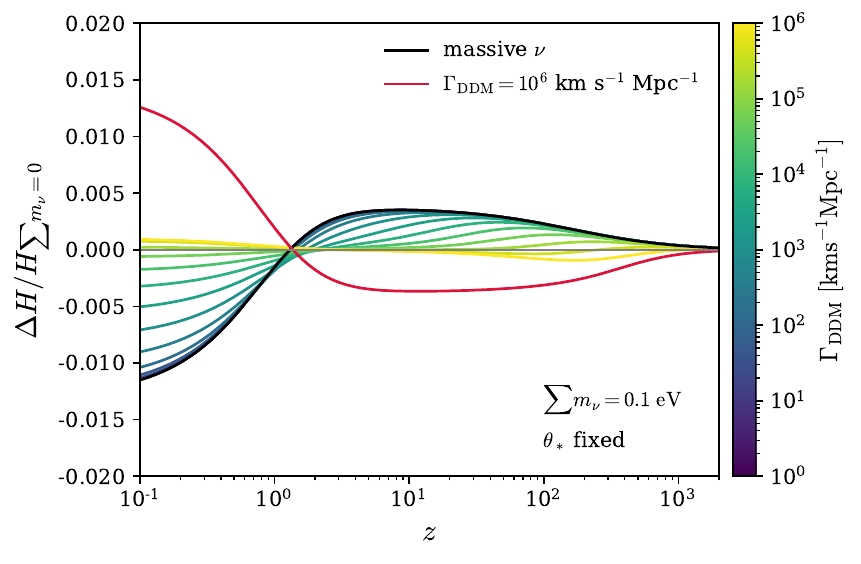}
    \includegraphics[width=0.49\linewidth]{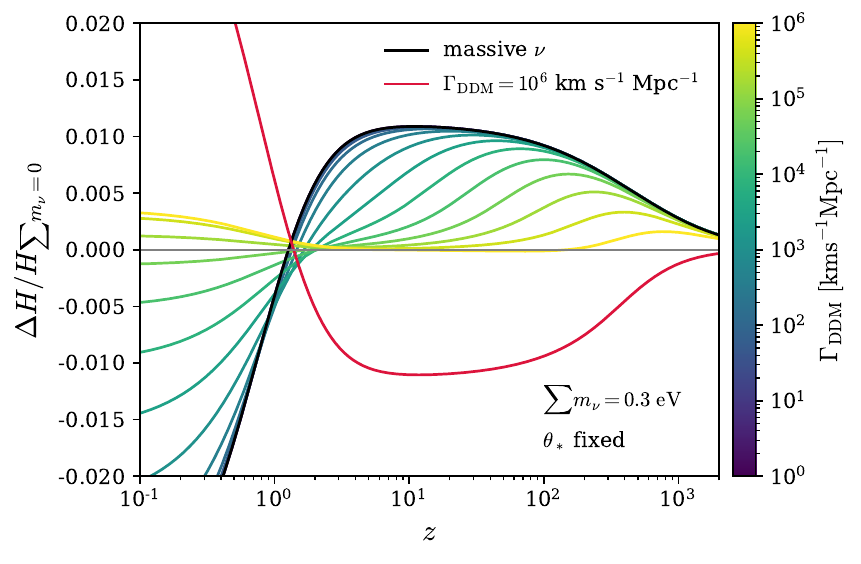}
    \caption{Fractional change to the Hubble rate in a cosmology with massive neutrinos  and/or DDM with massless products, normalized to that of a massless neutrino universe (similar to Fig.~1 of Refs.~\cite{Pan:2015bgi,Lynch:2025ine}). See main text for discussion.}
    \label{fig:deltaH_over_H}
\end{figure}

Within the DDM model, a degeneracy appears at the background level when the increase in matter density from neutrinos becoming non-relativistic is compensated by DM decays. This occurs when $\Omega^{\rm ini}_{\rm DDM} \approx \Omega_\nu$ and the decay rate satisfies $1/t_{\rm rec} \gg \Gamma \gg H_0$, where $t_{\rm rec}$ is the time at recombination. This is illustrated in Fig.~\ref{fig:deltaH_over_H}, where we show the fractional change to the Hubble rate in a cosmology with massive neutrinos and/or DDM with massless products, normalized to that of a massless neutrino universe (similar to Fig.~1 of Refs.~\cite{Pan:2015bgi,Lynch:2025ine}). We show $\sum m_\nu = 0.1$ eV on the left panel and  $\sum m_\nu = 0.3$ eV on the right panel, while the fraction of DDM is chosen such that $\Omega^{\rm ini}_{\rm DDM}=\Omega_\nu$. The angular acoustic scale at last-scattering $\theta_s$ is kept fixed by adjusting $H_0$ (automatically done via shooting in {\tt CLASS}). The color bar indicates different decay rates $\Gamma_{\rm DDM}\in\{0,10^6\}$ km/s/Mpc.  The chosen prior range ensures DM decay between recombination and today. The red curve assumes $\sum m_\nu =0$ to illustrate the sole effect of decaying dark matter. One can see that a short-lived DM fraction can compensate almost perfectly the effect of massive neutrinos at the background level. Larger neutrino masses require both a larger DDM fraction and a larger decay rate. Note, however, that a perfect match between the decay rate and the time of neutrino non-relativistic transition is not necessary to guarantee no observable signal. 

\subsection{Analysis setup and datasets}
\label{sec:analysis_setup}

To isolate the physical origin of neutrino mass constraints, we perform two distinct types of analyses. First, we carry out ``background-only'' analyses (Sec.~\ref{sec:background}) using combinations of BAO, SN1a, and CMB distance priors, which are sensitive primarily to the expansion history and geometric degeneracies. Second, we include the full CMB likelihood, allowing us to incorporate information from perturbations (Sec.~\ref{sec:cmb}), in particular CMB lensing, which probes the growth of structure and breaks degeneracies present at the background level.  

All cosmological analyses presented in this work are performed using either the public inference framework \texttt{Cobaya} \cite{Torrado:2020dgo} or \texttt{MontePython-v3} \cite{Audren:2012wb,Brinckmann:2018cvx}, which we employ for Bayesian parameter estimation via Markov Chain Monte Carlo (MCMC) sampling (performed with \texttt{mcmc} \cite{Lewis:2002ah} and analyzed with \texttt{GetDist} \cite{Lewis:2019xzd}). Convergence of the Bayesian posteriors is assessed using the Gelman--Rubin criterion $R-1<0.01$. In parallel, we perform a frequentist analysis and compute profile likelihoods for the sum of neutrino masses by maximizing the likelihood over all remaining parameters. For the background-only case, we use the algorithm \texttt{bobyqa} \cite{Powell2009bobyqa,Cartis2018,Cartis03082022} that is native to \texttt{Cobaya} for the minimization of the negative log-likelihood function, and \texttt{PROSPECT} \cite{Holm:2023uwa} to obtain the corresponding profile likelihood contours. In the full CMB case, we use \texttt{Procoli} \cite{Karwal:2024qpt}, a simulated-annealing optimizer, interfaced with \texttt{MontePython-v3}. Theoretical predictions for background quantities, linear perturbations, and cosmological observables are computed with the Boltzmann solver \texttt{CLASS} \cite{Blas:2011rf}. Analyses and plotting are performed in a \texttt{Python} framework consisting of the following packages: \texttt{JupyterLab} \cite{kluyver2016jupyter}, \texttt{Matplotlib} \cite{Hunter:2007}, \texttt{NumPy} \cite{harris2020array}, \texttt{pandas} \cite{mckinney-proc-scipy-2010} and \texttt{SciPy} \cite{2020SciPy-NMeth}.

We consider four classes of cosmological models throughout this work: the standard $\Lambda$CDM model with stable cold dark matter; a $\Lambda$DDM model in which a fraction of the dark matter decays into dark radiation; and extensions of both models allowing for a DDE equation of state parameterized by $(w_0,w_a)$. In all cases we assume three massive neutrino species with degenerate masses, unless stated otherwise, and impose flat priors on the cosmological parameters over ranges broad enough to avoid prior-induced constraints. The default prior ranges used in this work are given in Tab.\ \ref{tab:priors}. Prior ranges imposed in the DDM model ensure that DM decays as well as the non-relativistic transition of neutrinos happen well after recombination, as we want to restrict the analysis to late-time modifications of the expansion history. This ensures also that the CMB prior adopted for the background-only analysis applies. 

\begin{table}[]
    \centering
    \begin{tabular}{|c|c|c|}
        \hline 
         parameter&\multicolumn{2}{c|}{prior range}\\
          & Background-only &  Full-CMB  \\
         \hline 
         $H_{0}$ [km/s/Mpc] &  [50, 80] & [50, 80] \\
         $\omega_{b}$ & [0.005, 0.1] & [0.018, 0.03]    \\
         $\sum m_{\nu}$ [eV] & [0, 0.9] & [0, 0.9] \\
         \hline
         $\omega_\mathrm{c,tot}$ & [0.001, 0.99] & [0.1, 0.2] \\
         $f_{\mathrm{DDM}}$ & [$10^{-9}$, 0.1] & [$10^{-7}$, 0.1] \\
         $\log_{10} \Gamma$ & [3, 6] & [3, 6]\\
         \hline
         $\omega_{c}$ & [0.001, 0.99] & [0.09, 0.18]  \\
         $w_0$ & [-3, 1] &  [-3, 1]  \\
         $w_a$ & [-5, 2] & [-3, 2]   \\
         \hline
         $10^{9}A_s$ & N/A & [1.8, 3]   \\
         $n_s$ &  N/A & [0.9, 1]   \\
         $\tau_{\rm reio}$ & N/A & [0.004,0.12]   \\
         \hline
    \end{tabular}
    \caption{Default prior ranges used for the analyses of this work. Densities are given as $\omega_x = \Omega_x h^2$, with $\Omega_x$ the density parameter of component $x$ today relative to the critical density and $h$ being the Hubble constant in units of 100~km/s/Mpc. $\omega_{c,\rm tot}$ denotes the total dark matter density including the stable and the decaying component (evaluated today, assuming that most of DM has not decayed).}
    \label{tab:priors}
\end{table}

\vspace{0.5cm}
\noindent Our analysis relies on the following data:
\begin{itemize}
    \item \textbf{CMB prior}: For the background-only analyses, we employ compressed CMB information in the form of a Gaussian prior on the acoustic angular scale $\theta_*$, the baryon density $\omega_b$ and the CDM+baryon density $\omega_{bc}$. We use the values on $\theta_{*}$, $\omega_{b}$, $\omega_{bc}$ and their covariance provided in App.~A of \cite{DESI:2025zgx} which have been determined by \cite{Rosenberg:2022sdy} (using the NPIPE \texttt{CamSpec} CMB likelihood).

    \item \textbf{CMB data}: For the full CMB analyses, we use the \textit{Planck} 2018 (PR3) low-$\ell$ temperature and polarization Plik likelihood \cite{Aghanim:2019ame} and the NPIPE (PR4) \texttt{Camspec} likelihood for the high-$\ell$ TT, TE, and EE spectra \cite{Rosenberg:2022sdy}. Where specified, we also include the \textit{Planck} 2018 (PR3) CMB lensing likelihood \cite{Aghanim:2018oex} to probe the growth of structure.
    
    \item \textbf{BAO data}: We use the baryon acoustic oscillation measurements from the DESI DR2 release \cite{DESI:2025zgx}, including both isotropic and anisotropic distance constraints over the redshift range probed by the survey.
    
    \item \textbf{Type Ia supernovae}: We include uncalibrated distance modulus from the Pantheon$+$ compilation \cite{Scolnic:2021amr, Brout:2022vxf}, which provide complementary constraints on the late-time expansion history.
\end{itemize}

\noindent Finally, when reporting constraints, we give 2-sided error bars at 68\% C.L. and one side-errors at $95\%$ C.L.

\begin{table}[]
    \centering
    \resizebox{\textwidth}{!}{
    \begin{tabular}{l c | c c c c}
    \hline
    model/dataset & $\sum m_{\nu}$ [eV] & $\Omega_{m}$ & $\log_{10}\Gamma$ or $w_{0}$ & $f_{\mathrm{DDM}}$ or $w_{a}$ & $\Delta \chi^{2}_{\rm best\,fit}$\\
    \hline
    \bf{$\mathbf{\Lambda}$CDM} & & & & & \\
    BAO+CMB prior & $< 0.099$ [0.000] & $0.300\pm0.004$ [0.299] & $-$ & $-$ & $-$ \\
    SN1a+CMB prior & $0.228^{+0.084}_{-0.200}$ [0.168] & $0.336^{+0.013}_{-0.018}$ [0.331] & $-$ & $-$ & $-$ \\
    BAO+SN1a+CMB prior & $<0.109$ [0.002] & $0.301\pm 0.004$ [0.300] & $-$ & $-$ & $-$ \\
    \hline
    \bf{$\Lambda$DDM} &&&&& \\
    BAO+CMB prior & $\times$ [0.118] & $0.292^{+0.006}_{-0.005}$ [0.295] & $\times$ [5.36] & $>0.013$ [0.024] & $-3.6$ \\ 
    SN1a+CMB prior & $0.50^{+0.30}_{-0.20}$ [0.360] & $0.328\pm 0.017$ [0.332] & $\times$ [5.35] & $\times$ [0.015] & $-0.0$ \\
    BAO+SN1a+CMB prior & $\times$ [0.211] & $0.295\pm 0.005$ [0.297] &  $> 3.61 $ [5.99] & $>0.007$ [0.030] & $-3.0$\\
    \hline
    \bf{$w_{0}w_{a}$CDM} &&&&&\\
    BAO+CMB prior & $< 0.236$ [0.020] & $0.362^{+0.024}_{-0.031}$ [0.351] & $-0.32^{+0.24}_{-0.31}$ [$-0.43$] & $-2.08^{+0.99}_{-0.71}$ [$-1.68$] & $-6.7$ \\
    SN1a+CMB prior &  $\times$ [0.021] & $0.345\pm 0.021$ [0.327] & $-0.94\pm 0.12 $ [$-0.94$] & $-0.68^{+0.74}_{-0.59}$ [$-0.31$] & $-0.4$  \\
    BAO+SN1a+CMB prior & $< 0.158$ [0.002] & $0.311\pm 0.006$ [0.310] & $-0.85\pm 0.06 $ [$ -0.86$] & $-0.57\pm 0.23$ [$-0.43$] & $-7.1$ \\
    \hline
    \bf{$w_{0}$DDM} &&&&&\\
    BAO+CMB prior & $\times$ [0.188] & $0.297\pm 0.008$ [0.293] & $4.64^{+1.20}_{-0.62}$ [3.60] & $> 0.022$ [0.077] & $-5.0$\\
    &  &  & $-0.95\pm 0.05$ [$-0.91$] & $-$ & \\
    \cline{4-5}&  &  & &  &  \\
    SN1a+CMB prior &  $\times$ [0.459] & $0.317^{+0.024}_{-0.027}$ [0.295] &  $\times$ [5.40] & $\times$ [0.077] & $-0.4$\\
    &  &  & $-0.96^{+0.07}_{-0.06}$ [$-0.91$] & $-$ &  \\
    \cline{4-5}&  &  & &  &  \\
    BAO+SN1a+CMB prior & $\times$ [0.001] & $0.299^{+0.007}_{-0.006}$ [0.289] & $4.58\pm 0.80$ [3.06] & $> 0.026$ [0.081] & $-8.8$\\
    &  &  & $-0.93\pm 0.03$ [$-0.90$] & $-$ & \\
    \hline
    \bf{$w_{0}w_{a}$DDM} &&&&&\\
     BAO+CMB prior & $\times$ [0.189] & $0.338^{+0.026}_{-0.032}$ [0.349] & $\times$ [5.99] & $0.046^{+0.024}_{-0.034}$ [0.017] &  $-6.8$\\
    & & & $-0.54^{+0.24}_{-0.32}$ [$-0.48$] & $-1.30^{+1.00}_{-0.74} $ [$-1.48$]  & \\
    \cline{4-5}&  &  & &  &  \\
    SN1a+CMB prior & $\times$ [0.038] & $0.319\pm 0.026$ [0.276] &  $\times$ [6.00] & $\times$ [0.072] & $-0.4$ \\
    &  &  &  $-0.92^{+0.12}_{-0.11}$ [$-0.88$] &  $-0.32^{+0.66}_{-0.58}$ [$0.07$] &  \\
    \cline{4-5}&  &  & &  &  \\
    BAO+SN1a+CMB prior & $\times$ [0.047] & $0.303^{+0.008}_{-0.007}$ [0.305] & $\times$ [4.43] & $> 0.014 $ [0.021] & $-8.9$ \\
    &  &  & $-0.88\pm 0.06$ [$-0.89$] & $-0.24^{+0.28}_{-0.25}$ [$-0.24$] & \\
    \end{tabular}
    }
    \caption{Reconstructed mean and errors of cosmological parameters in various `background-only' analyses of the cosmology models under study. We give 2-sided error bars at 68\% C.L. and one sided limits at $95\%$ C.L. The values in brackets denote the corresponding best fit value at the likelihood maximum with $\Delta \chi^{2}$ values given with respect to the $\chi^{2}$ value of the maximum likelihood of $\Lambda$CDM. $\Omega_m$ is the matter density today (including dark matter, baryons and neutrinos). Crosses indicate that the corresponding constraint would be prior-limited and is therefore not displayed. For cases of non-zero $f_{\rm DDM}$, we give bounds at 95\% C.L.\ where possible.}
    \label{tab:background}
\end{table}


\section{Background-only cosmological analysis}\label{sec:background}

In this section, we focus only on cosmological datasets sensitive to the background expansion history, specifically BAO, type Ia supernovae, and CMB acoustic scale measurements. This approach allows us to isolate the impact of dark sector modifications on the expansion history, and quantify how much they can relax neutrino mass constraints independently of the impact of perturbations.


\subsection{Impact for neutrino mass constraints}

In Tab.~\ref{tab:background} we report constraints on $\sum m_\nu$ and the model parameters $\{\log_{10}\Gamma, f_{\rm DDM}\}$ and $\{w_0,w_a\}$ for the data set combinations BAO+CMB prior, SN1a+CMB prior and BAO+SN1a +CMB prior.
First and foremost, one can see as anticipated that constraints on the sum of neutrino masses are strongly relaxed in the extended cosmologies: they evolve from $\sum m_\nu < 0.109\,$eV under $\Lambda$CDM to $ \sum m_\nu  <0.158\,$eV in the DDE model to being fully unconstrained in the DDM model, i.e.\ compatible with our prior. The DDM model thus exhibits the largest degeneracy with neutrino masses at the background level.

\begin{figure}
    \centering
       \includegraphics[width=1\linewidth]{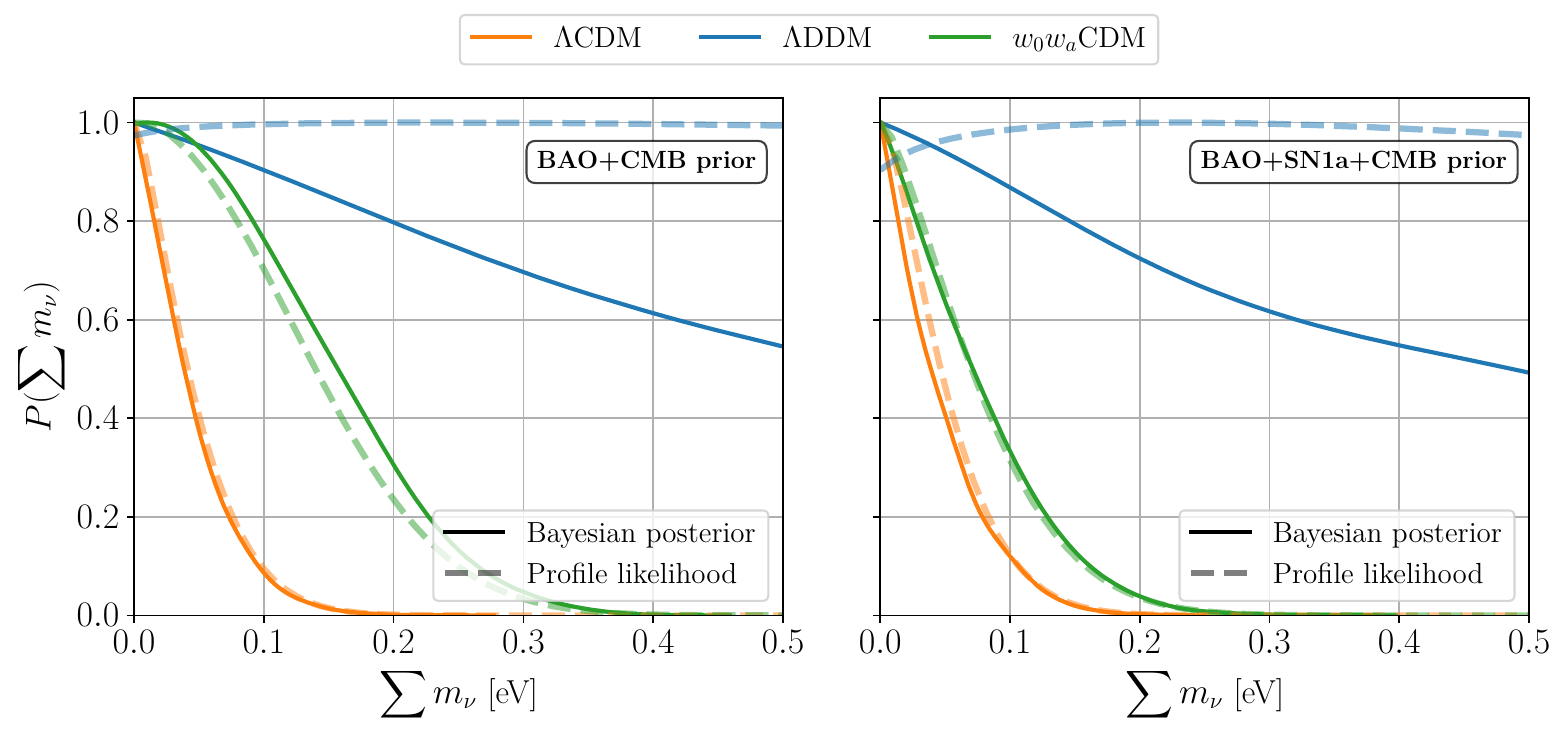}
       \includegraphics[width=1\linewidth]{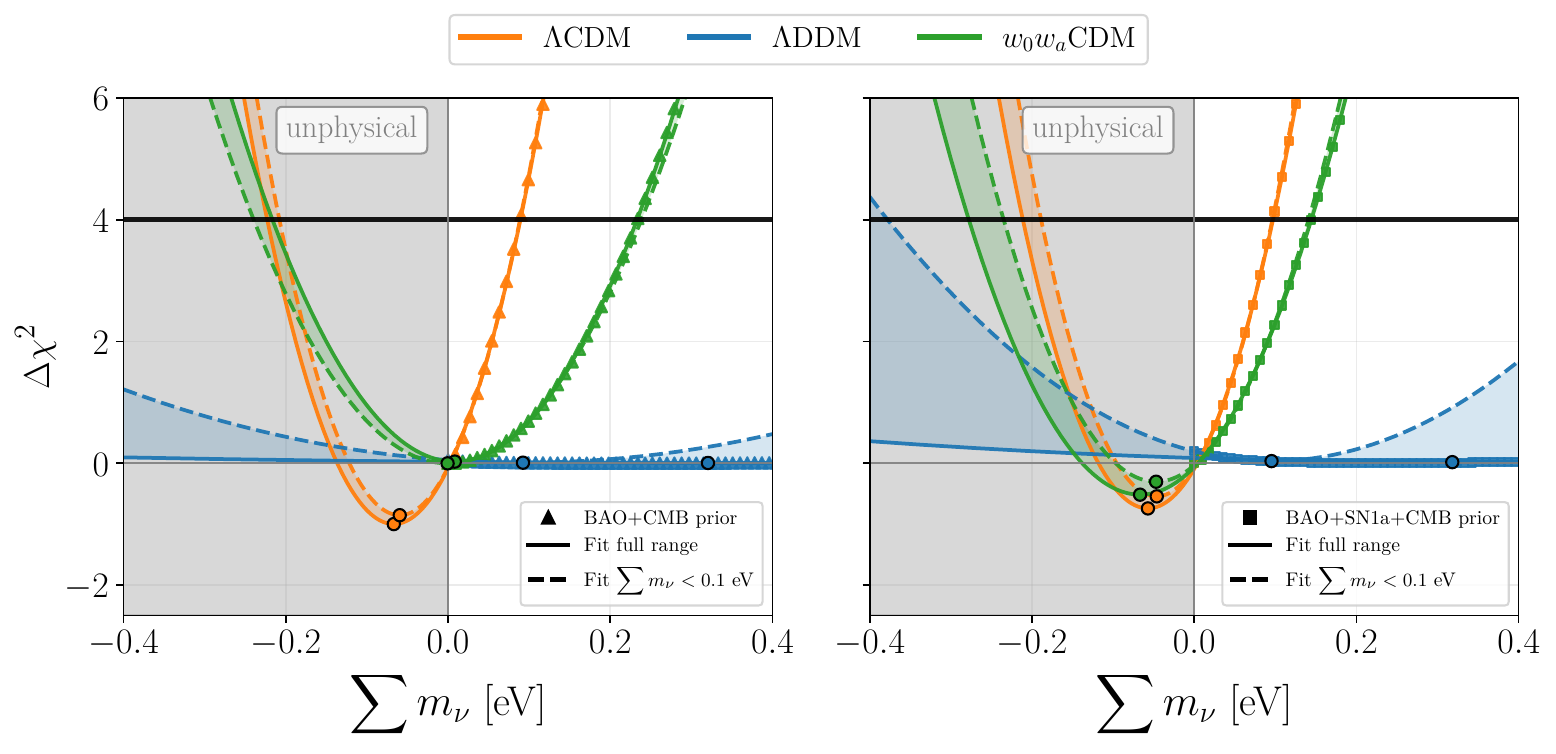}
       \caption{Bayesian posteriors and profile likelihoods of the neutrino mass sum for the cosmological models under study in the ``background-only'' analysis. Curves are given for BAO data with knowledge of CMB data (left) and when information from SN1a (in our case Pantheon+) is incorporated as well (right). The bottom plots compare the corresponding $\Delta\chi^{2}$ contours with parabola fits to indicate the best fit position of the neutrino mass sum. Here, filled regions indicate the effect of the chosen fit range, i.e.\ $\sum m_{\nu}<0.1\,$eV vs.\ $\sum m_{\nu}<0.9\,$eV.}
    \label{fig:mnu_bkg_profvspost}
\end{figure}

In Fig.~\ref{fig:mnu_bkg_profvspost}, top panels, we compare the posterior distribution (solid lines) and the profile likelihood (dashed line) of the neutrino mass sum in the various cosmological models for the data combinations BAO+CMB prior and BAO+SN1a+CMB prior. 
One can see that there is no constraint to the neutrino mass sum in the DDM model when considering solely background data. Upper limits in the Bayesian framework appear only due to volume effects (and would change with a different prior choice), whereas they are entirely relaxed in the frequentist analysis as apparent by the flat profile likelihood or $\Delta\chi^2$ curves in Fig.~\ref{fig:mnu_bkg_profvspost}. Note that volume effects are negligible for $\Lambda$CDM and $w_0w_a$CDM; in these cases Bayesian posteriors and profile likelihood curves match very well. 

In the bottom panel of Fig.~\ref{fig:mnu_bkg_profvspost}, we show parabolic fits to the corresponding $\Delta \chi^{2}$ curves to indicate which models prefer a negative sum of neutrino masses. We find again that $\Lambda$CDM favors negative neutrino masses regardless of SN1a inclusion \cite{Elbers:2025vlz}. In contrast, positive values near zero are favored in the $w_0w_a$CDM model in the BAO+CMB prior analysis, but shift to slightly negative values when SN1a is included. In all cases, negative best fit points have $\Delta\chi^2 \gtrsim -1$ with respect to $\sum m_\nu = 0$ and are therefore not significant in this analysis. In the DDM cosmology $\Delta\chi^2$ is nearly flat for both datasets, reflecting the discussed degeneracy.

\begin{figure}
    \centering
     \includegraphics[scale=0.38]{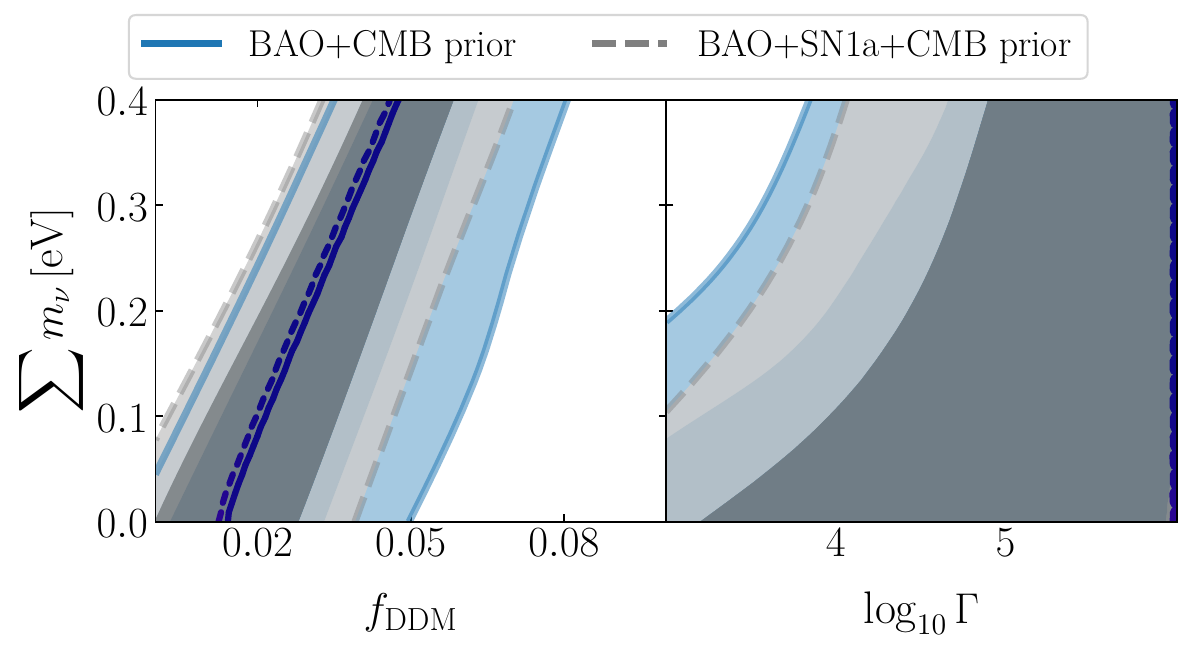}
     \includegraphics[scale=0.38]{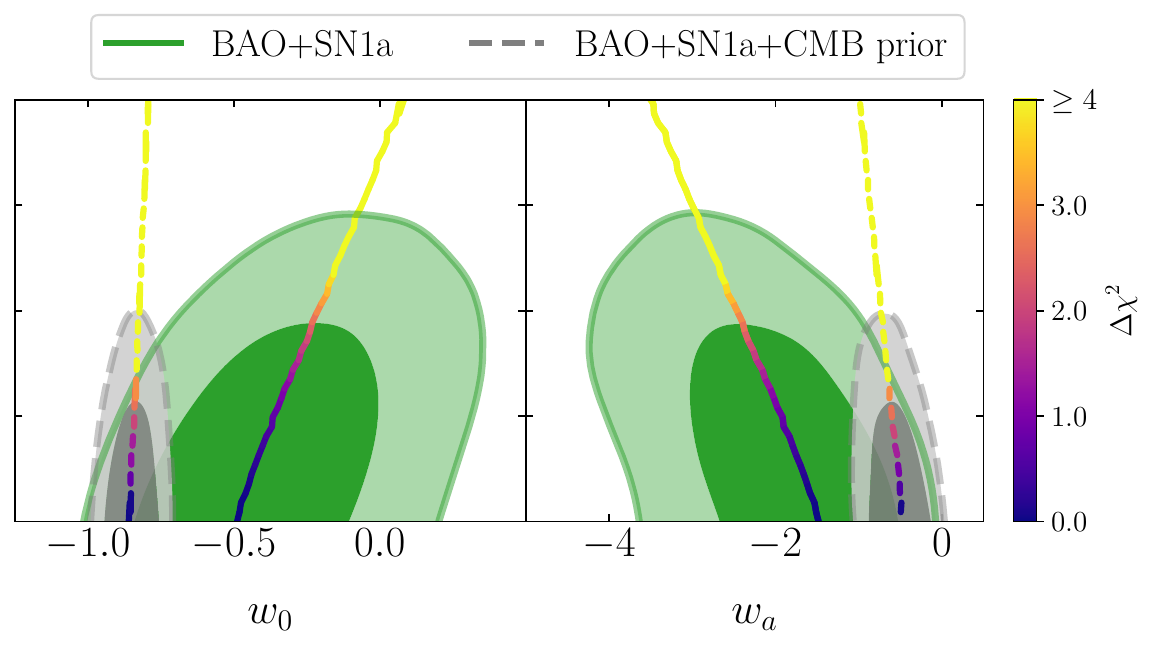}
    \caption{Constraints on the sum of neutrino masses as a function of dark sector parameters in the background-only analysis. Left panels: Decaying Cold Dark Matter ($\Lambda$DDM) model showing the fraction of decaying dark matter ($f_{\rm DDM}$) and decay rate ($\log_{10}\Gamma$) vs.\ $\sum m_\nu$. Right panels: Dynamical dark energy model ($w_{0}w_{a}$CDM) showing $(w_0, w_a)$ vs.\ $\sum m_\nu$.  Colored lines indicate the best fit profile likelihood paths with $\Delta\chi^2$ encoded with color. The color bar is truncated at $\Delta\chi^2 = 4$ for clarity.}
    \label{fig:bkg_post}
\end{figure}

\begin{figure}
    \centering
    \includegraphics[trim=0 0 0 0,clip, width=\linewidth]{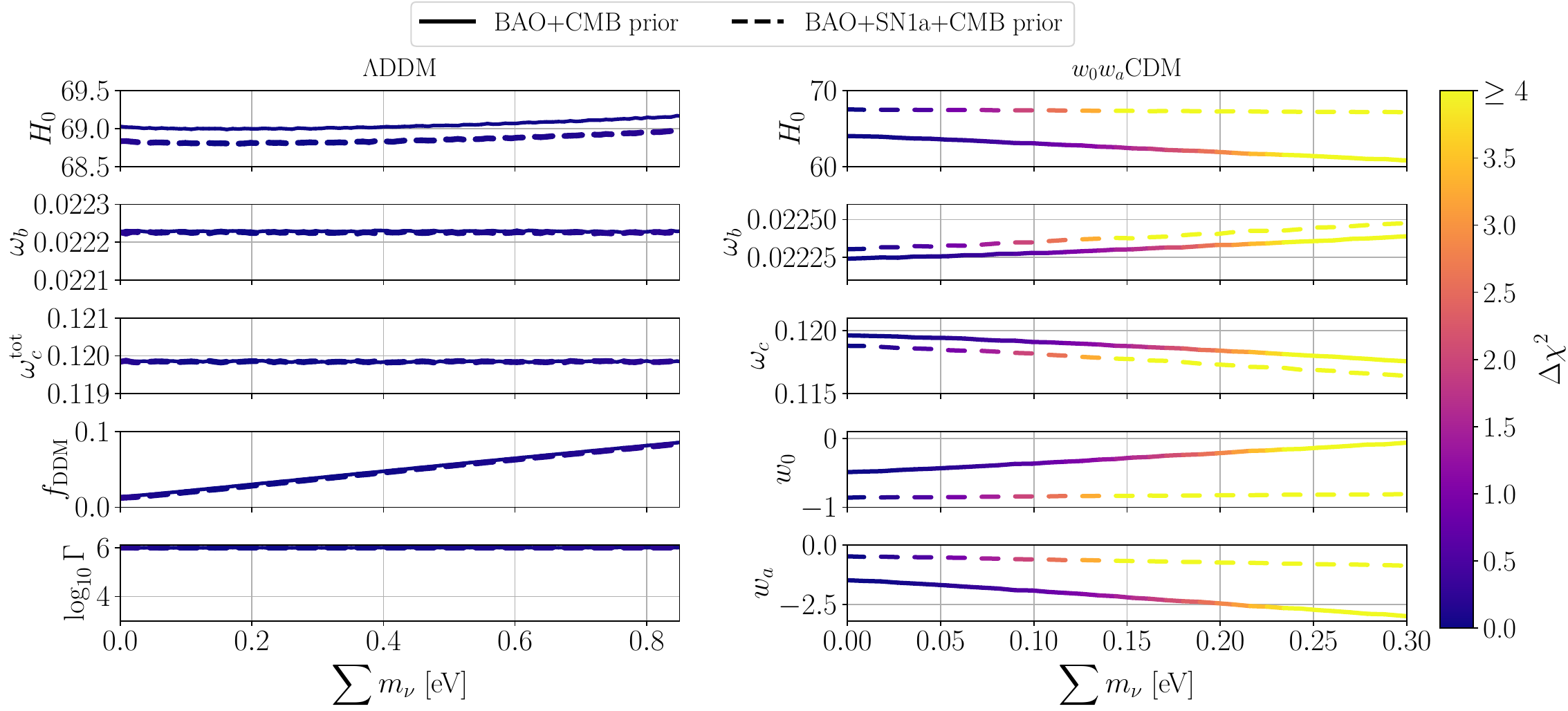} 
    \caption{Parameter evolution as a function of the sum of neutrino masses (profiled likelihood) for the models $\Lambda$DDM (left) and $w_{0}w_{a}$CDM (right), respectively, in the background-only analysis. The $\Delta\chi^2$ along the paths is color-coded.
    Results for BAO+CMB prior are indicated with solid lines, while results including data from SN1a are shown with dashed lines. The degeneracy between $f_{\mathrm{DDM}}$ and $\sum m_{\nu}$ is clearly visible in the left plots. $\omega_c^{\rm tot}$ corresponds to the dark matter density including the decaying fraction. In contrast, larger neutrino masses are compensated by an interplay of multiple parameters in the $w_0w_a$CDM model shown in the right plot.}
    \label{fig:profiled_likelihood}
\end{figure}

We further illustrate the degeneracy between $\sum m_\nu$ and other cosmological parameters in the posterior distribution in Fig.~\ref{fig:bkg_post} and in the profile likelihood shown in Fig.~\ref{fig:profiled_likelihood}, color-coded by the $\Delta \chi^2$ with respect to the best fit model. 

For $\Lambda$DDM (left panel), we recover the strong degeneracy between $\sum m_\nu$ and $f_{\rm DDM}$ that was expected ($\omega^{\rm ini}_{\rm DDM}\sim \omega_\nu = \sum m_\nu / 93.14~(\rm {eV})$), keeping $\Delta\chi^2$ close to $0$ even at large neutrino masses, while leaving the other parameters unaffected. Interestingly, the posteriors demonstrate that it is not necessary that the decay rate matches the time of non-relativistic transition of the neutrinos. Rather, it is enough that the decay rate be large enough for the DDM decay to cancel out the effect of neutrino masses within the redshift range probed by BAO and SN1a data.  Consequently, the lower limit to $\log_{10}\Gamma$ increases with the neutrino mass sum. 
The upper limit instead is prior dominated and independent of the neutrino mass the best fit always converges to the highest decay rate $\Gamma$ allowed by our prior, though in practice the likelihood is nearly flat within the prior range considered. Note also the small change to $H_0$ necessary to keep the CMB acoustic scale fixed. The inclusion of Pantheon+ data does not break the degeneracy between $\sum m_{\nu}$ and $f_{\mathrm{DDM}}$, and the profile with and without Pantheon+ data are identical.

The DDE model (right panels) shows a similar but much weaker degeneracy between the neutrino mass and the dark energy parameters $w_0$ and $w_a$. While both Bayesian and frequentist analyses agree on the degeneracy direction, the compensation mechanism is far less effective than for DDM: high neutrino masses significantly degrade the fit quality above $\sum m_\nu>0.2$~eV. Moreover, the inclusion of supernovae data further degrades the correlation between the neutrino mass sum and the CPL parameters $(w_{0}, w_{a})$, as visible both in the profile and in the MCMC results.

\begin{figure}
    \centering
    \includegraphics[width=0.7\linewidth]{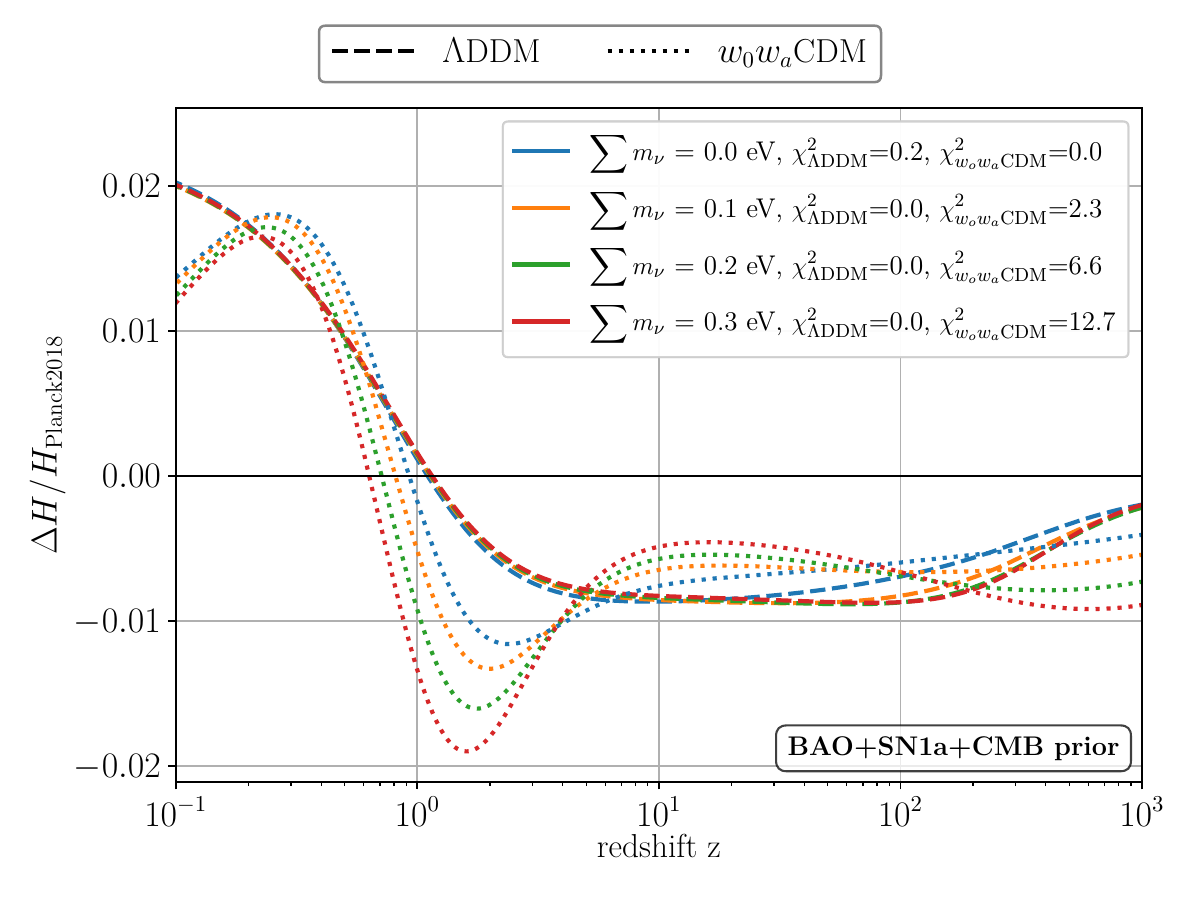}
    \caption{Change of the Hubble parameter along the neutrino mass profiles for $\Lambda$DDM and $w_{0}w_{a}$CDM using the combined BAO+SNa1+CMB prior data. For $\sum m_\nu = (0.0, 0.1, 0.2, 0.3)$~eV, $\Delta \chi^{2}$ values of (0.2, 0.0, 0.0, 0.0) for $\Lambda$DDM and (0.0, 2.3, 6.6, 12.7) for $w_{0}w_{a}$CDM are obtained with respect to the models' individual best fit values. Best fit parameters of {\it Planck} 2018 TTTEEE+lensing are used as reference.}
    \label{fig:Delta_Hz}
\end{figure}

Finally, we illustrate the degeneracy directly at the level of the expansion history in Fig.~\ref{fig:Delta_Hz}, where we plot the deviation to the expansion history (normalized to the {\it Planck} 2018 $\Lambda$CDM prediction) at various points along the neutrino mass profiles. This is interesting in several ways: 
First it further confirms that, under DDM, all curves are degenerate even for large neutrino masses. This is not the case in the DDE model, where deviation appears at virtually all redshifts and the degeneracy with neutrino masses is only partial. 
Second, one can see that there is a non-zero deviation from $\Lambda$CDM detected in both DDE and DDM. While the former is expected from the previously reported hint for DDE in recent analyses of the same datasets, the latter is novel and suggests an alternative explanation to the ``DESI tension''. We further explore this in the next subsection.

\begin{figure}
    \centering
    \includegraphics[trim=5 5 5 5,clip,width=\linewidth]{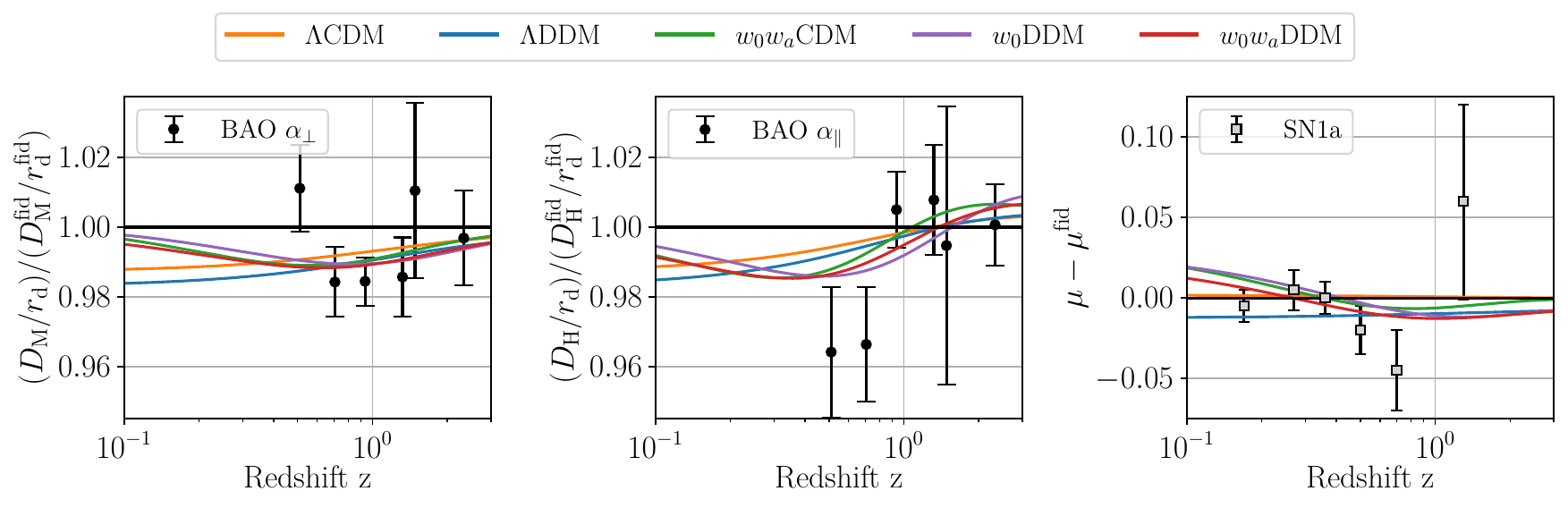}
    \caption{BAO scaling parameters and residual distance modulus for the models under study. Their predictions for parameters at maximal likelihood of BAO+SN1a+CMB prior are shown together with data of DESI and Pantheon+. {\it Planck} 2018 TTTEEE+lensing is used as reference. Data points are taken from Ref.~\cite{DESI:2025zgx}.
    }
    \label{fig:distances}
\end{figure}
\begin{figure}
    \centering
    \includegraphics[width=\linewidth]{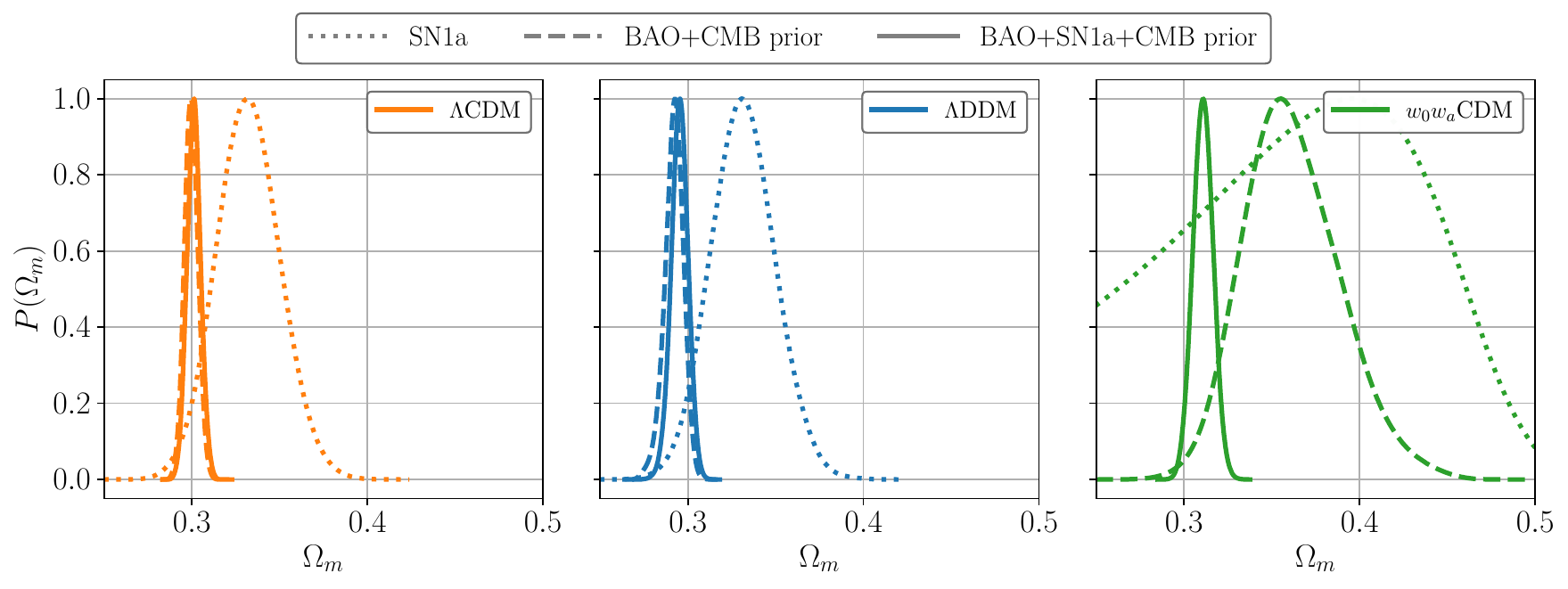}
    \caption{Posteriors of $\Omega_{m}$ for $\Lambda$CDM (left), $\Lambda$DDM (middle), $w_{0}w_{a}$CDM (right) under consideration of different datasets.}
    \label{fig:Omega_m}
\end{figure}
\begin{figure}
    \centering
    \includegraphics[trim=10 15 10 10,clip,width=\linewidth]{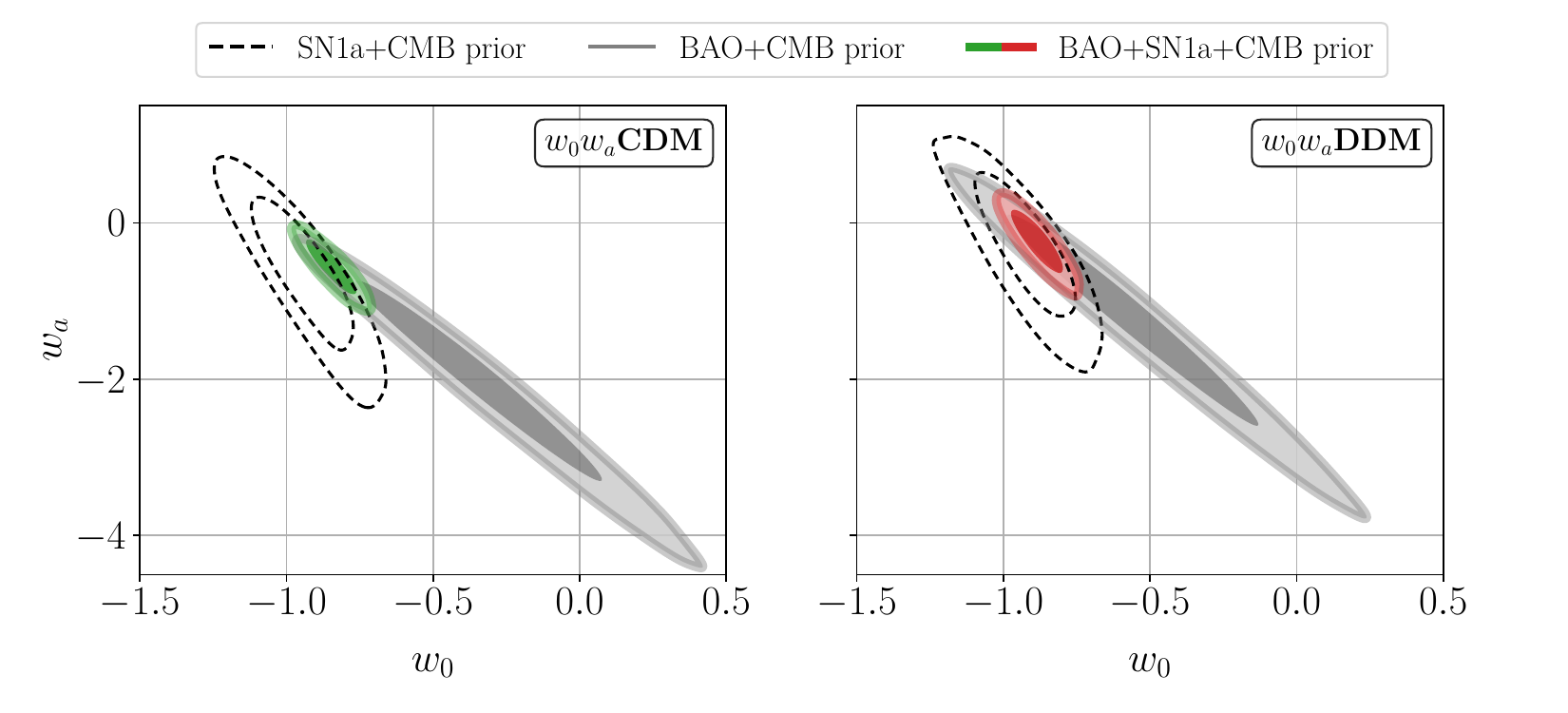}
    \caption{Comparison of the $w_{0}$-$w_{a}$ contour between $w_{0}w_{a}$CDM and $w_{0}w_{a}$DDM for different cosmological datasets. Note that for the combination of BAO and supernova data with the applied CMB prior, $w_{0}w_{a}$CDM favors $w_{a}\leq$0, while in $w_{0}w_{a}$DDM $w_{a}=0$ is compatible with data.}
    \label{fig:w0wa_extended}
\end{figure}

\begin{figure}
    \centering
    \includegraphics[width=\linewidth]{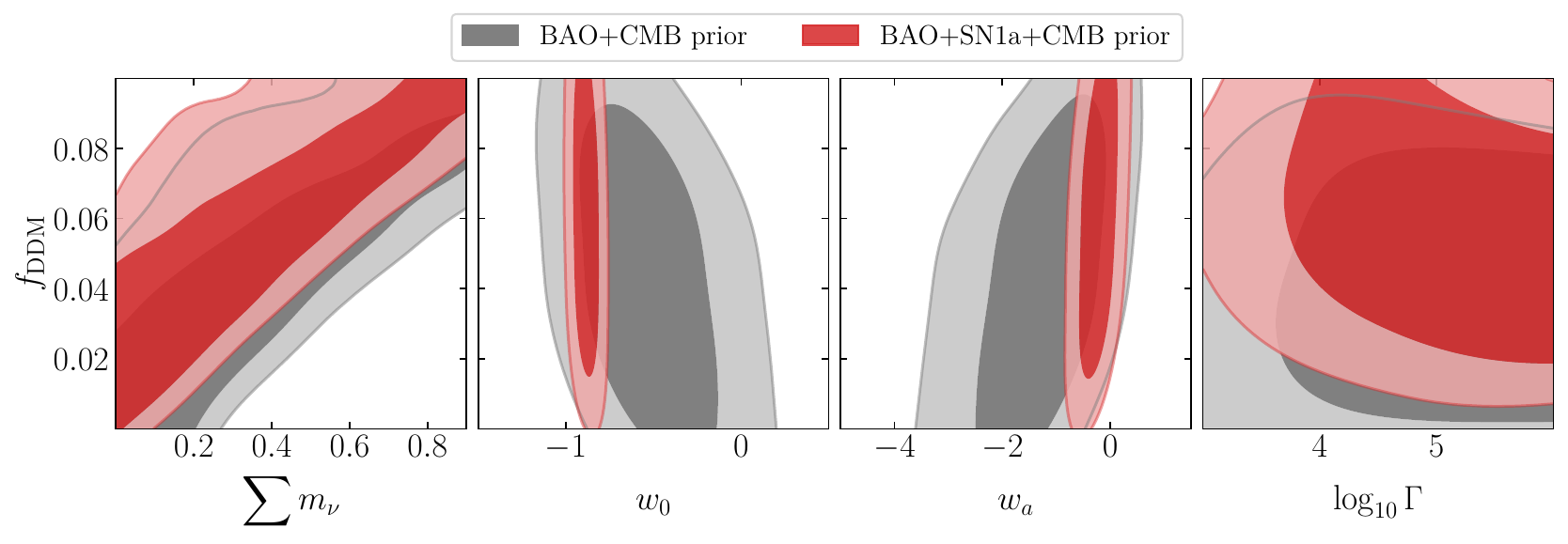} \\
    \includegraphics[width=\linewidth]{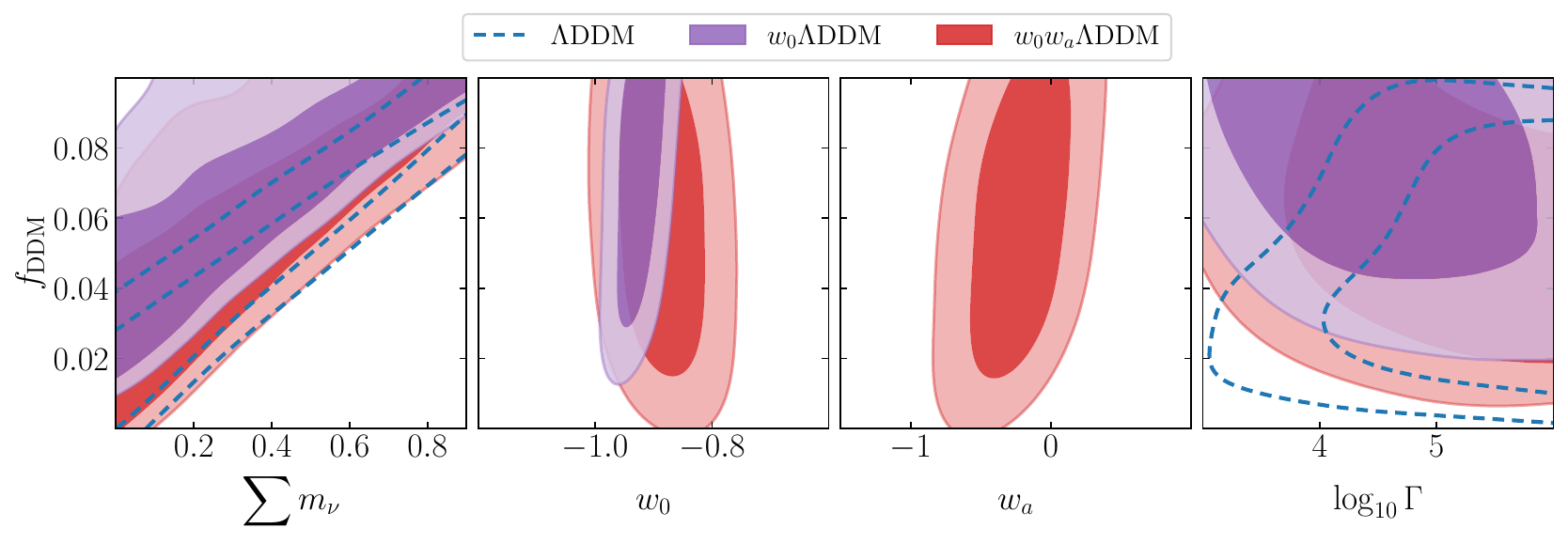}
    \caption{Fraction of DDM $f_{\mathrm{DDM}}$ in dependence of sum of neutrino masses $\sum m_{\nu}$, the DE equation of state parameters $w_{0,a}$ and the DM decay rate $\log_{10}\Gamma$ in the background-only analysis. The impact of including supernova information is shown in the top panels for the $w_0w_a$DDM model, while the contour regions of different cosmological models for BAO+SN1a+CMB prior data is illustrated in the bottom panels.} 
    \label{fig:ddm_fraction}
\end{figure}


\subsection{Preference for dark sector physics}

The preference for DDE we observe is consistent with those reported in Ref.~\cite{DESI:2025zgx} by the DESI collaboration. Although the potential of DDM to explain the DESI tension was mentioned in Ref.~\cite{Lynch:2025ine}, it was not explicitly quantified. We find here (see Tab.~\ref{tab:background}) that the background data combination of BAO+CMB prior yields $f_{\rm DDM}>0.013$ ($\log_{10} \Gamma$ is unconstrained within our prior choice) with $\Delta\chi^2 = -3.6$ in favor of $\Lambda$DDM over $\Lambda$CDM. The inclusion of SN1a data however slightly reduces the preference for a non-zero DDM fraction, yielding $f_{\rm DDM}>0.007$ and $\log_{10}\Gamma >3.61$, with $\Delta\chi^2= -3.0$. Note that the absence of upper-limit to $f_{\rm DDM}$ is due to the tight correlation with $\sum m_\nu$. 

To understand the (small) preference for $\Lambda$DDM, we show in Fig.~\ref{fig:distances} the perpendicular ($\alpha_\perp$) and parallel ($\alpha_\parallel$) BAO, as well as the SN distance modulus $\mu$, predicted in the DDM and DDE models and normalized to the {\it Planck} $\Lambda$CDM best fit. Data points of DESI and Pantheon+ are also displayed. One can see that the $\Lambda$DDM solution is similar to the $\Lambda$CDM fit to DESI alone, with a low $\Omega_m$ value. However, as $\Omega_m$ has decreased over time, it can provide a better fit to the CMB prior by having a larger $\omega_{\rm CDM}^{\rm tot}=\omega_{\rm DDM}+\omega_{\rm CDM}$ at early times. Note that DDE appears to do a better job at adjusting the low$-z$ SN1a (and slightly the parallel BAO points). This is expected given that our model effectively behaves like a low-$\Omega_m$ solution at those redshifts, which cannot describe well SN1a data.

This is further illustrated in the comparison between the reconstructed $\Omega_m$ posteriors shown in Fig.~\ref{fig:Omega_m}. Compared to $\Lambda$CDM, the $\Lambda$DDM model broadens and shifts low the $\Omega_m$ posteriors, improving the consistency between the datasets. However, it does not perform as well as $w_0w_a$CDM, that shifts up the DESI $\Omega_m$ posterior in agreement with Pantheon+. Note that the combined analysis in the CPL model favors a $\Omega_m$ value in the tail of the individual distributions, indicating a complicated degeneracy breaking with $w_0-w_a$ in the higher-dimensional posterior.

\begin{figure}
    \centering
    \includegraphics[width=\linewidth]{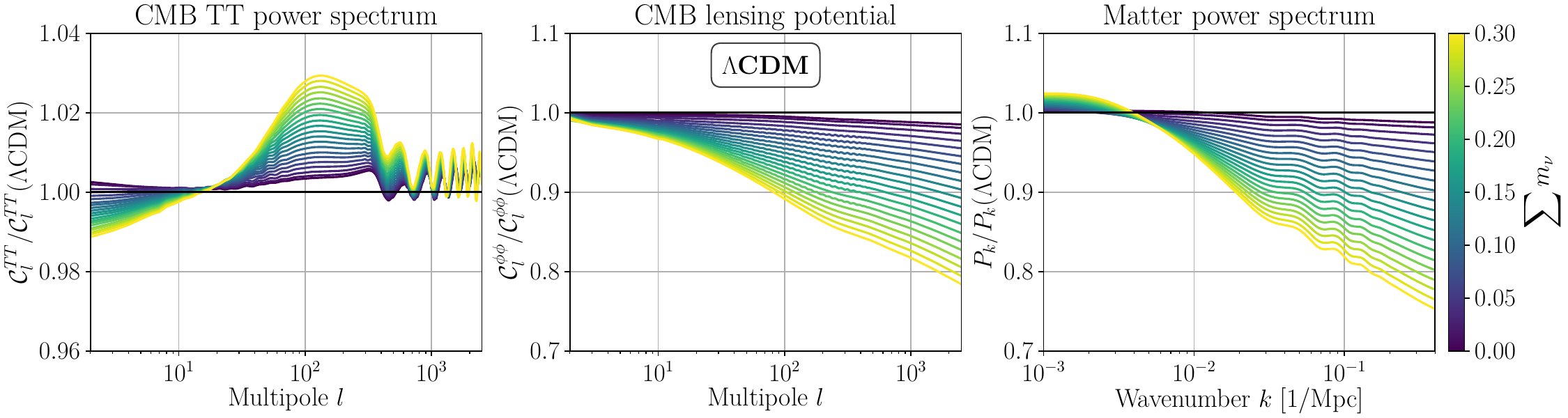} \\
    \includegraphics[width=\linewidth]{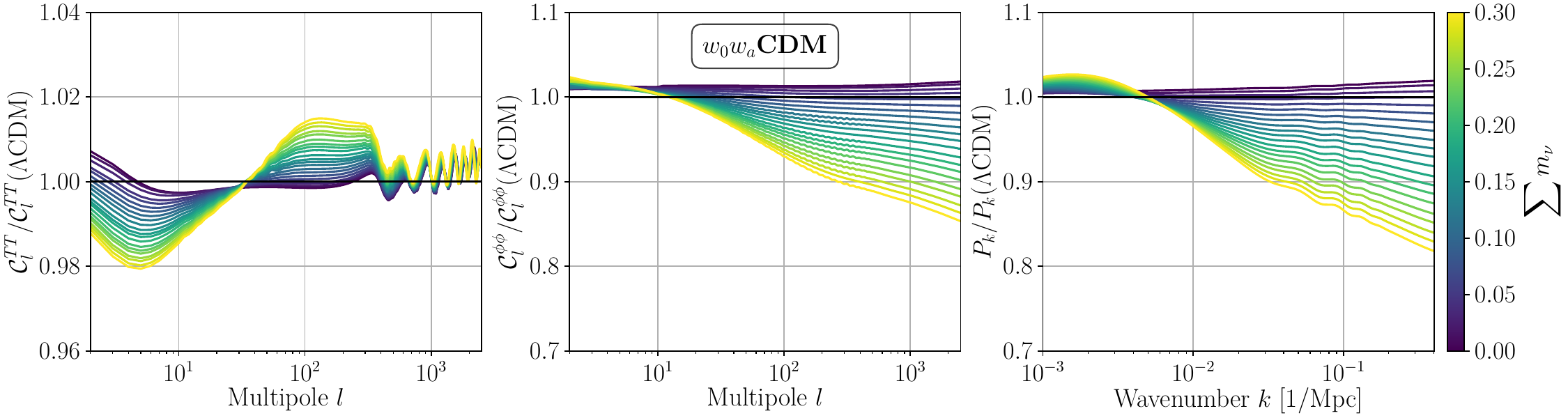}\\
    \includegraphics[width=\linewidth]{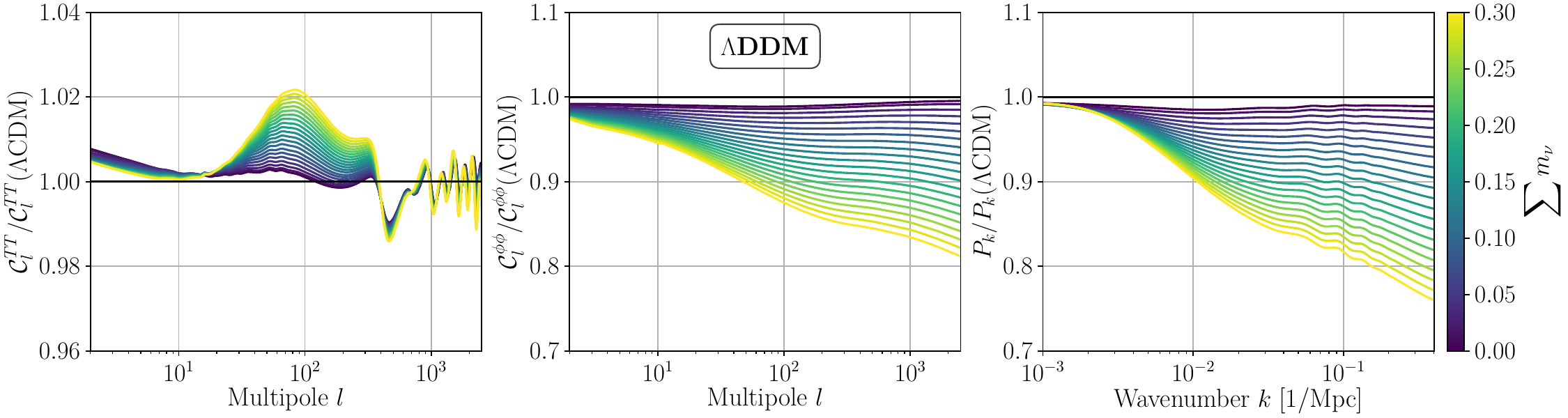} 
    \caption{Impact of neutrino mass on the CMB angular temperature power spectrum $\mathcal{C}_{l}^{TT}$ (left), the CMB lensing potential power spectrum $\mathcal{C}_l^{\phi\phi}$ (middle) and the matter power spectrum $P_{k}$ (right) for the models $\Lambda$CDM (top panels), $w_{0}w_{a}$CDM (middle panels) and $\Lambda$DDM (bottom panels), respectively. The underlying parameters are taken from the profiled background-only likelihood of BAO+SN1a+CMB prior. Here, $A_{s}$, $n_{s}$ and $\tau_{\mathrm{reio}}$ are set to the best fit values of our BAO+CMB+SN1a analysis.}
    \label{fig:power_spectra_numasses}
\end{figure}

Finally, we study the possibility of simultaneously allowing for DDE and DDM, i.e.\ $w_0$DDM and $w_0w_a$DDM. We provide reconstructed parameters and best fit values in Tab.~\ref{tab:background} and show the associated predictions (parameters at maximal likelihood) for BAO and supernova distances in Fig.~\ref{fig:distances}. As previously mentioned, we expect the effect of the DM decay on the expansion history to reproduce that of the phantom era. This is already partially visible in Fig.~\ref{fig:Delta_Hz}, where $H(z)$ appears to increase at low-$z$ due to the larger value of $\Omega_{\Lambda} =1-\Omega_m$, that is a consequence of a smaller $\Omega_m$ after the DM decay. We show in Fig.~\ref{fig:w0wa_extended} the Bayesian contours of $\{w_0,w_a\}$ in the $w_0w_a$CDM and $w_0w_a$DDM models from the BAO+SN1a+CMB~prior analysis. Importantly, while $w_a=0$ is excluded at $\sim 2\sigma$ in $w_0w_a$CDM, adding DDM ($w_0w_a$DDM) shifts $w_a$ toward zero, making a constant dark energy equation of state compatible with the data.

Correlations of $f_{\mathrm{DDM}}$ and the neutrino mass as well as the other dark sector parameters are illustrated in Fig.~\ref{fig:ddm_fraction}. The top panels compare different datasets in the $w_0w_a$DDM model, while the bottom panels compare different models ($\Lambda$DDM, $w_0w_a$DDM and $w_0$DDM) using the BAO+SN1a+CMB prior dataset. One can see that $f_{\rm DDM}> 0$ and $w_0 > -1$ are favored at $\sim 2\sigma$ when the datasets are combined. 

In fact, we further test that restricting $w_a=0$, i.e.\ running solely with a constant $w_0$, yields an equally good fit to BAO+SN1a+CMB prior as the full $w_0w_a$DDM model, with $w_0 > -1$ at $\sim 2\sigma$ (see bottom panel of Fig.~\ref{fig:ddm_fraction}). We also note that the degeneracy between $f_{\rm DDM}$ and $\{w_0,w_a\}$ weakens the detection of those parameters compared to the case where these extensions are considered separately. However, the addition of SN1a data strengthens their detection, in particular that of $f_{\rm DDM}$. This suggests a potential alternative explanation to the `CMB-DESI tension', in which the phantom behavior of DE is replaced by a decrease in the CDM density (due to the unstable component in our model), while DE behaves as a simple quintessence field. Some other scenarios that couple dark matter and dark energy \cite{Wolf:2024stt,Teixeira:2024qmw,Cataneo:2025vae,Smith:2025grk,Khoury:2025txd} also have a similar phenomenological behavior at the background level. A discussion of how dark forces affect structure growth in this context can be found in \cite{Costa:2025kwt}.

\noindent We show in the next section how including the effect of perturbations alters these conclusions.

\begin{table}
    \centering
    \resizebox{\textwidth}{!}{
    \begin{tabular}{l c | c c c c}
    \hline
     model/dataset & $\sum m_{\nu}$ [eV] & $\Omega_{m}$ & $\log_{10}\Gamma$ or $w_{0}$ & $f_{\mathrm{DDM}}$ or $w_{a}$ & $\Delta \chi^{2}_{\rm best\,fit}$\\
    \hline
    \bf{$\mathbf{\Lambda}$CDM} & & & & & \\
    BAO+CMB+SN1a  & $< 0.080$ [0.000] & $0.302\pm 0.004$ [0.301] & $-$ & $-$ & $-$  \\
    \quad ... + lensing & $< 0.071$ [0.000] & $0.302\pm 0.004$ [0.302] & $-$ & $-$ & $-$ \\
    \hline
    \bf{$\Lambda$DDM} &&&&& \\
    BAO+CMB+SN1a  & $< 0.130$ [0.000] & $0.300\pm 0.004$ [0.300] & $4.62^{+1.30}_{-0.53}$ [5.98] & $< 0.025$ [0.006] & $-0.7$  \\
    \quad ... + lensing & $< 0.079$ [0.002] & $0.300\pm 0.004$ [0.300] & $\times$ [ 5.93] & $< 0.019$ [0.006] & $-0.5$ \\
    \hline
    \bf{$w_{0}w_{a}$CDM} &&&&&\\
    BAO+CMB+SN1a  & $< 0.145$ [0.000] & $0.312\pm 0.006$ [0.311] & $-0.84\pm 0.06$ [-0.85] & $-0.59^{+0.23}_{-0.20}$ [-0.50] & $-7.8$  \\
    \quad ... + lensing & $< 0.125$ [0.002] & $0.311^{+0.005}_{-0.006}$ [0.311] & $0.84\pm 0.06$ [-0.85] & $-0.60^{+0.24}_{-0.20}$ [-0.51]& $-8.4$ \\
    \hline
    \bf{$w_{0}$DDM} &&&&&\\
    BAO+CMB+SN1a  & $< 0.111$ [0.000] & $0.303^{+0.005}_{-0.006}$ [0.305] & $\times$ [5.76] & $0.018^{+0.007}_{-0.014} $ [0.012] & $-3.4$  \\
     &  &  &$-0.96\pm 0.03$ [-0.96] & $-$ &  \\
    \cline{4-5}&  &  &  &  &  \\
     \quad ... + lensing & $<0.070$ [0.000] & $0.304\pm0.005$ [0.304] & $\times$ [5.80] & $0.012^{+0.004}_{-0.011}$ [0.009] & $-2.1$ \\ 
     &  &  & $-0.97^{+0.03}_{-0.02}$ [-0.97] & $-$ & \\ 
    \hline
    \bf{$w_{0}w_{a}$DDM} &&&&&\\
    BAO+CMB+SN1a  & $< 0.166$ [0.001]  & $0.310\pm 0.006$ [0.310]  & $\times$ [4.85] & $< 0.026$ [0.000] & $-7.8$  \\
    &  &  & $-0.85\pm 0.11$ [$-0.85$] & $-0.49^{+0.43}_{-0.45}$ [$-0.49$] &  \\
    \cline{4-5}&  &  & &  &  \\
     \quad ... + lensing & $<0.135$[0.001] & $0.310\pm 0.006$ [0.310] & $\times$ [4.95] & $< 0.019$ [0.000] & $-8.4$ \\
     & & & $-0.85\pm 0.06$ [-0.85] & $-0.55^{+0.24}_{-0.22}$ [-0.52] & 
    \end{tabular}
    }
    \caption{Reconstructed mean and errors of cosmological parameters in various `full CMB' analyses of the cosmology models under study. We give 2-sided error bars at 68\% C.L. and one sided limits at $95\%$ C.L. The values in brackets denote the corresponding best fit value at the likelihood maximum with $\Delta \chi^{2}$ values given with respect to the $\chi^{2}$ value of the maximum likelihood of $\Lambda$CDM. Crosses indicate that the corresponding constraint would be prior-limited and is therefore not displayed.}
    \label{tab:cmb}
\end{table}


\section{Constraints from the full datasets}\label{sec:cmb}


In this section, we turn to including the full CMB power spectra likelihoods to illustrate the role played by perturbations in constraining the DDM scenarios and the sum of neutrino masses, and in particular the role of CMB lensing. On the other hand, we show that constraints in the DDE model are not significantly affected by the inclusion of perturbations, when DDE is modeled as a smooth non-clustering field.


\subsection{The impact of perturbations}

To gauge the role of perturbations in constraining models, we show in Fig.~\ref{fig:power_spectra_numasses} the residuals of the CMB temperature, lensing and matter power spectra (normalized to our $\Lambda$CDM best fit of BAO+CMB+SN1a) when following the background degeneracy between $\sum m_\nu$ and the extended models, uncovered through the profile likelihood of the `background-only' data.
We fix the values of the primordial power spectrum amplitude $A_{s}$, spectral index $n_{s}$ and optical depth $\tau_{\mathrm{reio}}$ to the best fit values found in our analysis of BAO+CMB+SN1a and generate power spectra with the best fit parameters along the likelihood profile of $\sum m_\nu$ in the background-only analysis, i.e.\  $\{\omega_b, \omega_{bc}, H_0\}$ and either $\{w_0,w_a\}$ or $\{f_{\rm DDM},\log_{10}\Gamma\}$. Each curve is color-coded according to the value of the sum of neutrino masses. 

The visible effects are typical of neutrino masses as reviewed (e.g.) in Refs.~\cite{Lesgourgues:2006nd,Lesgourgues:2018ncw}, with a power suppression particularly visible at small-scales in the CMB lensing and matter power spectrum. The effect of the suppressed CMB lensing power spectrum is also visible in the temperature spectrum at large multipoles, as small oscillations. The larger effects at intermediate multipoles in TT are due to changes to the integrated Sachs-Wolfe effect when neutrinos becomes non-relativistic close to recombination. However, they are significantly reduced in the  $w_0w_a$CDM cosmology compared to $\Lambda$CDM (roughly a factor of $\sim 2$ in all observables), except at very large scales in TT, where cosmic variance anyway dominates. Note in particular the power suppression in the matter and CMB lensing power spectra is less pronounced, with a small increase on large-scales (due to slightly different values of $\Omega_m$). In the $\Lambda$DDM model however, the suppression remains unaffected at small-scales, and slightly increases on large scales, due to the late-time DM decay.
As we show below, these features turn out to be particularly important in constraining neutrino masses in the DDM model, much more than in the case of DDE.

\begin{figure}
    \centering
    \includegraphics[scale=0.39]{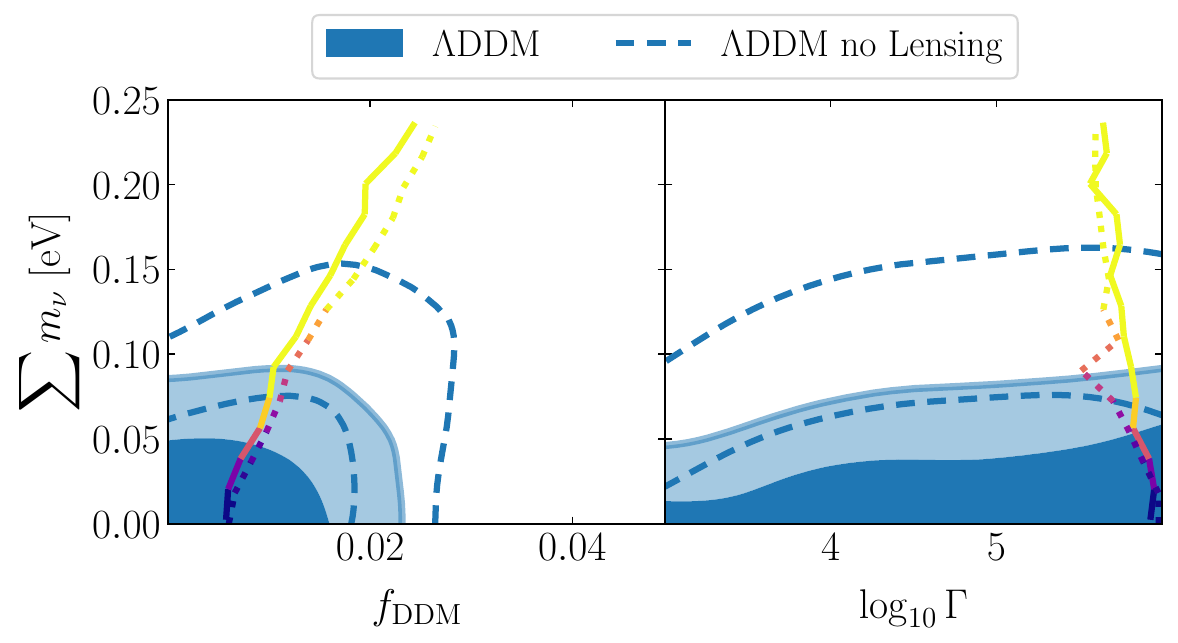}
    \includegraphics[scale=0.39]{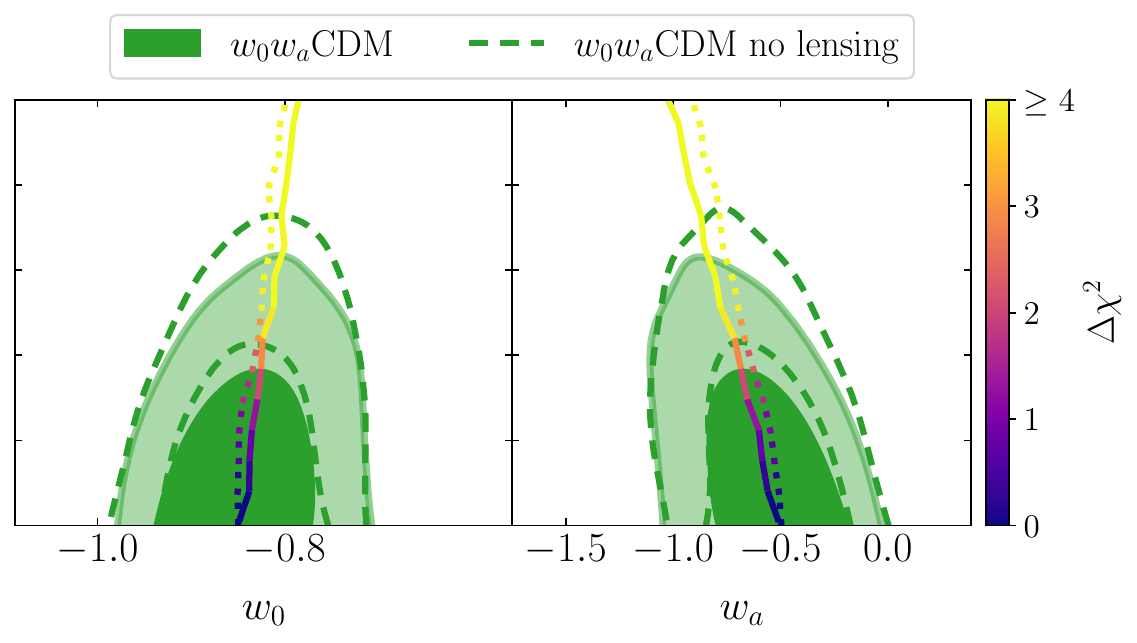}
    \caption{Constraints on the sum of neutrino masses as a function of dark sector parameters in the full CMB analysis. Left panels: Decaying Cold Dark Matter (DDM) model showing the DDM fraction ($f_{\rm DDM}$) and decay rate ($\log_{10}\Gamma$) vs. $\sum m_\nu$. Right panels: Dynamical dark energy model ($w_0w_a$CDM) showing  $(w_0, w_a)$ vs.\ $\sum m_\nu$. We are using BAO+CMB+SN1a data, where filled contours and solid lines represent constraints with lensing data included, while dashed contours and dotted lines show constraints without lensing. Colored lines indicate the best fit profile likelihood paths with $\Delta\chi^2$ encoded in color. The color bar is truncated at $\Delta\chi^2 = 4$ for clarity.}
    \label{fig:full_2D}
\end{figure}

\begin{figure}
    \centering
   \includegraphics[width=1\linewidth]{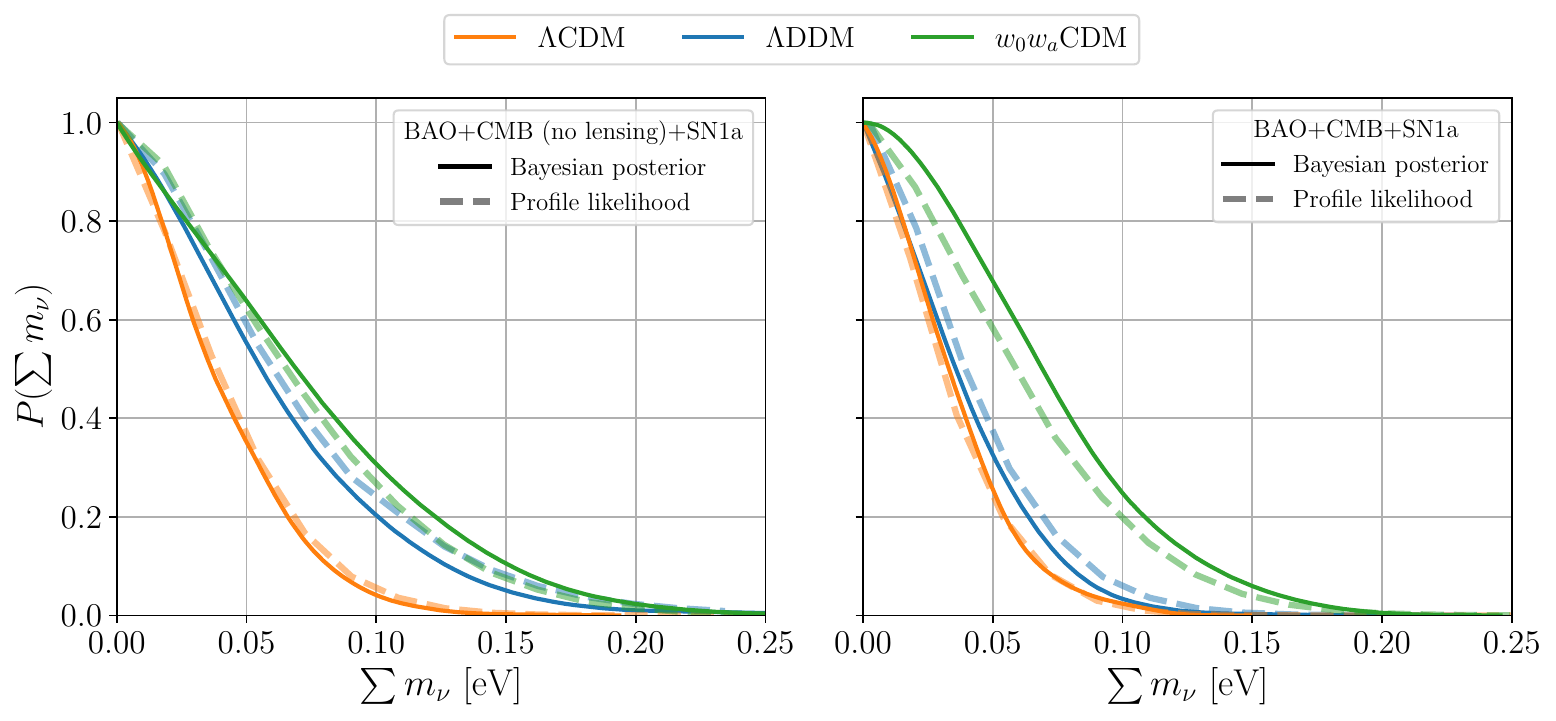}
   \includegraphics[width=1\linewidth]{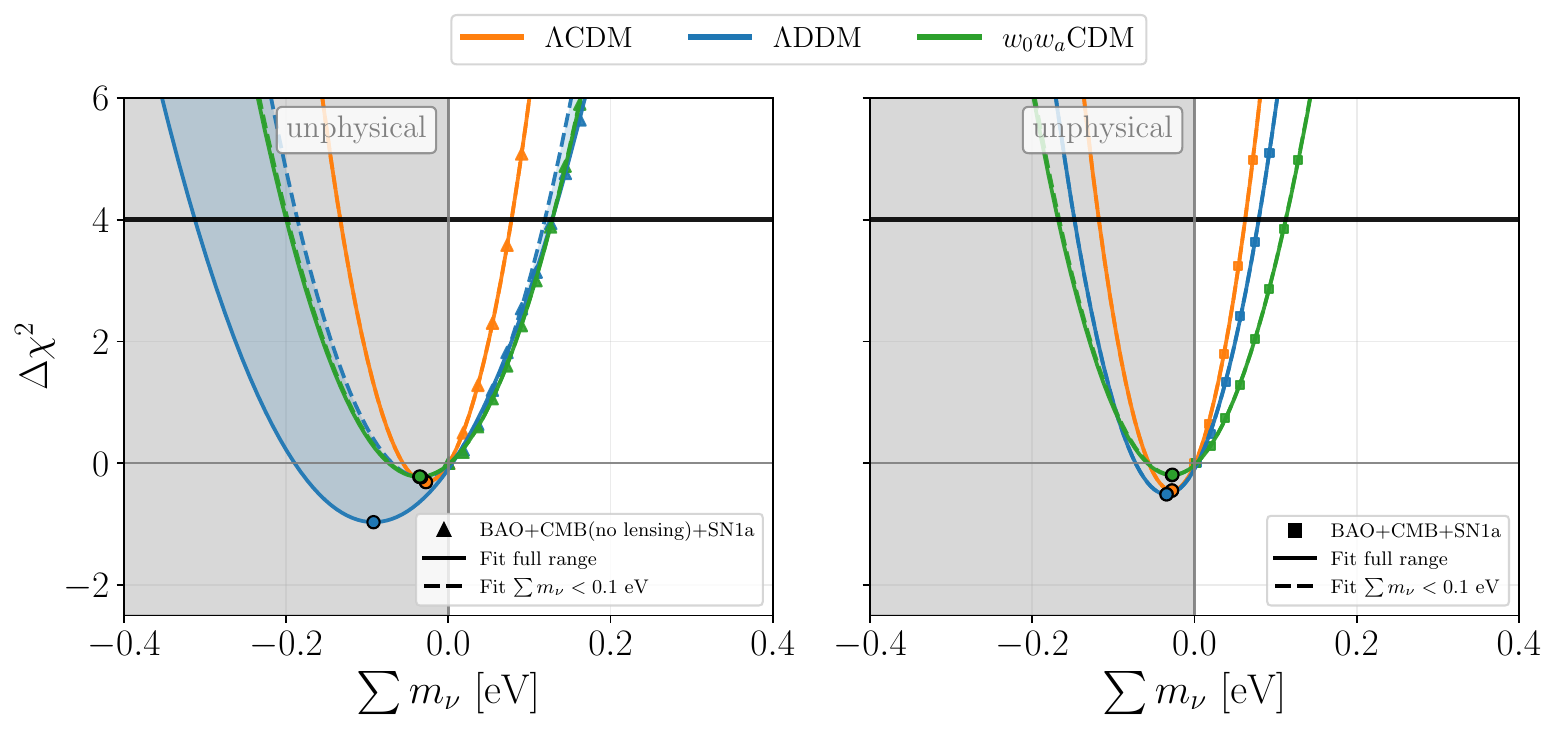}
    \caption{Bayesian posteriors and profile likelihoods of the neutrino mass sum for the cosmological models under study in the “full CMB” analysis. Curves are given for the combination of BAO+CMB+SN1a datasets. To underline the effect of lensing, we explicitly show results without lensing (left) and results including lensing (right) in the analyses of CMB data. The bottom plots illustrate the corresponding $\Delta\chi^{2}$ contours with parabola fits to the neutrino mass sum. In contrast to the background-only case, only $\Lambda$DDM (without lensing) is sensitive to the chosen fit range ($\sum m_{\nu}<0.1\,$eV vs.\ $\sum m_{\nu}<0.9\,$eV), i.e.\ shows deviations from quadratic shape.}
    \label{fig:full_1D}
\end{figure}


\subsection{Analyses with full CMB data}

We now compute the full constraints to $\sum m_\nu$ in the two extended cosmological scenarios, as well as the joint $w_0w_a$DDM and $w_0$DDM models. We perform two sets of analyses to gauge the impact of CMB lensing on the constraints. As previously, we check the robustness of our results to prior volume effects by also computing the profile likelihood of $\sum m_\nu$. Our results are summarized in Tab.~\ref{tab:cmb}, where we give the parameter constraints, and shown in Figs.~\ref{fig:full_2D} and \ref{fig:full_1D}.

\begin{figure}[]
    \centering
    \includegraphics[height=0.55\textheight]{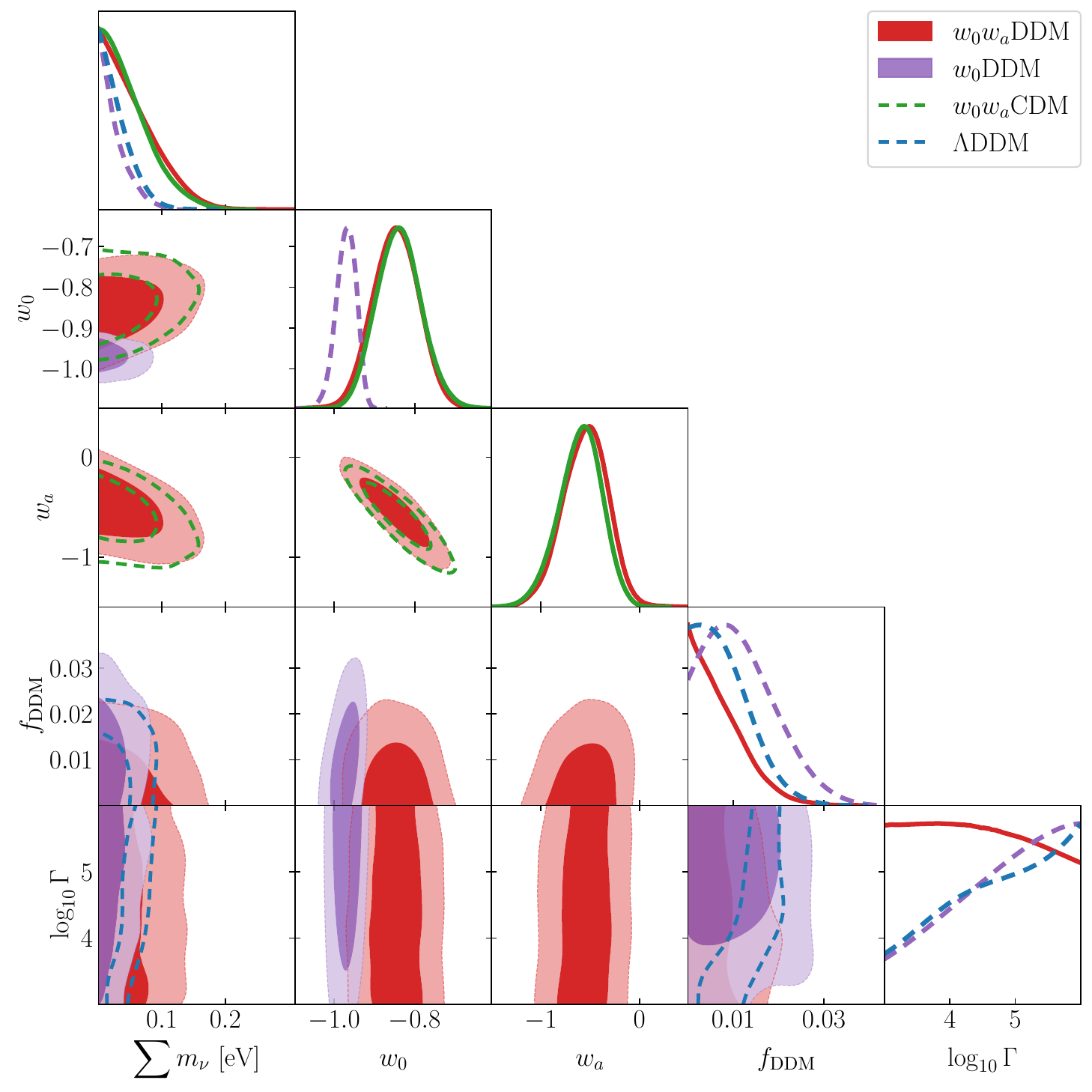}
    \caption{Contour plots of the cosmological models for the combined analysis of BAO, CMB and SN1a data. The indication for a non-zero DDM fraction $f_{\mathrm{DDM}}$ vanishes in this case, such that the contours of $w_0w_a$DDM and $w_0w_a$CDM match each other.}
    \label{fig:full_combined}
\end{figure}

First and foremost, one can see the major improvement of the neutrino mass constraints in the $\Lambda$DDM scenario, now restricted to $\sum m_\nu < 0.13$ eV (without lensing) and $\sum m_\nu < 0.079$ eV (with lensing). In comparison, constraints in the $w_0w_a$CDM model are fairly stable, $\sum m_\nu < 0.145$ eV (without lensing) and $\sum m_\nu < 0.125$ eV (with lensing). This is confirmed by the profile likelihood analyses in Fig.~\ref{fig:full_1D} that yield similar constraints and show no evidence of prior volume effects. Indeed, the effect of DDM, which suppresses power on small scales, adds to that caused by massive neutrinos rather than it compensates it. Further note that lensing has a strong impact on the $\Delta \chi^{2}$ profile of DDM as it reduces the preference for negative neutrino masses and the deviation from quadratic shape. Second, the previous hint of a detection of non-zero $f_{\rm DDM}$ disappears in the full analysis, while the parameters $\{w_0,w_a\}$ are left largely unaffected compared to the background-only analysis. The combination of DDM and DDE, $w_0w_a$DDM, shown in Fig.~\ref{fig:full_combined}, exhibits no more deviation from the standard DDE analysis. The degeneracy between the phantom behavior of DE and DDM is thus fully broken. This underscores the crucial role played by perturbations and growth of structures measurements: while models may be performing very well in exploiting background degeneracy with neutrino masses, and in adjusting the CMB/BAO angles, the inclusion of growth of structures measurements is able to firmly exclude models that lead to too strong power suppression.

\section{Conclusions}\label{sec:conclusions}


Recent neutrino mass bounds obtained when combining the latest LSS spectroscopic surveys with CMB observations are in tension with the lower limits implied by terrestrial experiments. Barring an unknown systematic effect or a statistical fluke, this suggests some physics beyond $\Lambda$CDM or the standard model of particle physics.

In this work, we have revisited cosmological constraints on the sum of neutrino masses in scenarios that relax the standard assumptions about the late-time dark sector. We focused in particular on two well-studied extensions of the $\Lambda$CDM model: a dynamical dark energy component described by the CPL parametrization ($w_0w_a$CDM), and a scenario in which a fraction of the cold dark matter decays into an invisible relativistic species on cosmological timescales ($\Lambda$DDM).
While the former scenario was already shown to significantly alter neutrino mass bounds, $\sum m_\nu < {\cal{O}}(0.2-0.3)$ eV, we have found that DDM exhibits an even stronger degeneracy with massive neutrinos at the background level.
Indeed, by compensating the late-time increase in the matter density induced by neutrinos becoming non-relativistic, the DDM scenario can render background-only datasets effectively insensitive to $\sum m_\nu$. A profile likelihood analysis reveals that values as large as $\mathcal{O}(1\,\mathrm{eV})$ are allowed without degrading the fit. 
In this sense, it provides a more efficient mechanism than DDE for evading background-based neutrino mass bounds.

In the background-only analysis based on a CMB prior combined with DESI BAO and Pantheon+ SN1a data, we also found a mild preference for a non-zero short-lived decaying dark-matter fraction, with $f_{\rm DDM}>0.007$ and $\log_{10}(\Gamma/[{\rm km/s/Mpc}]) >3.61$, suggesting an alternative explanation to the CMB-DESI tension that does not rely on phantom dark energy. Furthermore, when both DDM and DDE are allowed simultaneously, the decay of dark matter can mimic the phantom behavior usually inferred in CPL analyses, thereby alleviating the need for a crossing of the phantom divide while maintaining compatibility with supernova data through a quintessence-like equation of state.

However, this degeneracy is decisively broken once perturbation observables are included. While massive neutrinos and DDM have opposite effects on the background expansion, their impact on the growth of cosmic structures is cumulative: both suppress power on small scales. As a consequence, the inclusion of the full \textit{Planck} CMB data, and in particular CMB lensing, restores strong constraints on the neutrino mass in the DDM scenario. We find that the upper bound on $\sum m_\nu$ is only moderately weakened relative to $\Lambda$CDM, reaching $\sum m_\nu \lesssim 0.079\,\mathrm{eV}$ when lensing information is included. At the same time, the apparent preference for a non-zero decaying component disappears, and the fraction of decaying DM is constrained as $f_{\mathrm{DDM}}\lesssim 0.019$. Furthermore, the degeneracy between DDM and phantom dark energy is fully lifted; constraints in the CPL dark energy model are much less affected by the inclusion of perturbations, reflecting the fact that smooth dark energy primarily alters the expansion history rather than the growth of structure.

Taken together, our results highlight the crucial role of structure-growth observables in assessing extensions of the dark sector and cosmological neutrino mass bounds. While background-only constraints on the sum of neutrino masses can be highly permissive in the presence of non-standard dark sector physics, the combination of geometric and perturbation probes can be used to break degeneracies and obtain robust bounds to the sum of neutrino masses. Looking ahead, this suggests that viable alternatives to $\Lambda$CDM capable of relaxing neutrino mass bounds must reproduce the background signatures favored by DESI and SN1a data while leaving the growth of perturbations essentially unaltered compared to $\Lambda$CDM. Exploring such scenarios, and confronting them with upcoming Stage-IV large-scale-structure and CMB datasets, represents a promising direction for future work.


\begin{acknowledgments}
We thank Gabriel Lynch and Lloyd Knox for helpful comments on this manuscript.
TM and VP acknowledge the European Union’s Horizon Europe research and innovation programme under the Marie Skłodowska-Curie Staff Exchange grant agreement No 101086085 – ASYMMETRY.
This work received funding support from the European Research Council (ERC) under the European Union’s HORIZON-ERC-2022 (grant agreement no. 101076865).
We gratefully acknowledge support from the CNRS/IN2P3 Computing Center (Lyon - France) for providing computing and data-processing resources needed for this work.
Development of code and refinement of figures were assisted by the use of AI tools, i.e.\ Microsoft 365 Copilot.
\end{acknowledgments}


\bibliographystyle{JHEP}
\bibliography{biblio.bib}

\end{document}